\documentclass[a4paper,11pt]{article}
\usepackage{jheppub} 
\usepackage{lineno}
\usepackage[utf8]{inputenc}
\usepackage[english]{babel}
\usepackage{csquotes}
\usepackage{physics}
\usepackage[T1]{fontenc}
\usepackage{graphicx}
\usepackage{chngcntr}
\usepackage{subcaption}
\usepackage{tabularx}
\usepackage{caption}
\renewcommand{\figurename}{Figure}
\usepackage{hyperref}
\usepackage[colorinlistoftodos]{todonotes}
\usepackage{epigraph}
\usepackage[export]{adjustbox}
\usepackage{amsmath}
\allowdisplaybreaks
\usepackage{amsthm}
\usepackage{amssymb}
\usepackage{amscdx}
\usepackage{nccmath}
\usepackage{mathtools}
\usepackage{mathrsfs}
\usepackage{blindtext}
\usepackage{titlesec}
\usepackage{natbib}
\usepackage{comment}
\usepackage{simplewick}
\usepackage{dsfont}
\usepackage{cancel}

\newcommand{\parsc}{\!\bullet\!}
\newcommand{\trasc}{\!\circ\!}

\usepackage{subfiles}

\preprint{}

\title{Lightcone Bootstrap for Multipoint Defect Correlators}

\author[a,b]{Lorenzo Bianchi,}
\author[a,b]{Andrea Mattiello,}
\author[a,b]{Lorenzo Quintavalle}
 \affiliation[a]{Dipartimento di Fisica, Università di Torino,\\Via P. Giuria 1, 10125 Torino, Italy}
\affiliation[b]{INFN - Sezione di Torino,\\Via P. Giuria 1, 10125 Torino, Italy}

\emailAdd{lorenzo.bianchi@unito.it, andrea.mattiello@unito.it, lorenzo.quintavalle@unito.it}

\abstract{We initiate the lightcone bootstrap analysis of multipoint correlators in a defect conformal field theory. The setup we consider is the three-point function of two bulk and one defect operator. Requiring consistency of the crossing equation in the lightcone limit, we find constraints on the defect spectrum at large transverse spin. Specifically, to reproduce the exchange of the leading-twist operator in the bulk channel we find two twist-accumulating families of defect operators at large transverse spin and we compute their defect CFT data in this limit.}

\begin{document}
\maketitle
\flushbottom

\newpage

\section{Introduction and discussion}
\label{section:Introduction}
The central idea of the conformal bootstrap program is to constrain the dynamics of conformal field theories (CFTs) using symmetries and internal consistency. After the breakthrough of \cite{Rattazzi:2008pe}, this program was revived, leading to important numerical and analytical results (see \cite{Rychkov:2023wsd, Bissi:2022mrs} and references therein). One of the early achievements of the analytic bootstrap is due to \cite{Komargodski:2012ek,Fitzpatrick:2012yx}, where it was found that a generic CFT must contain an infinite tower of double-twist operators, whose dimension approaches the generalized free field (GFF) dimension at large spin. The main idea leading to this result was to exploit the crossing equation for the four-point function of identical operators in Lorentzian signature and consider its behaviour in the lightcone limit, i.e. when two operators become lightlike separated. Several important developments followed, either trying to make the statement mathematically rigorous \cite{Qiao:2017xif,Kravchuk:2020scc,Kravchuk:2021kwe,Pal:2022vqc,vanRees:2024xkb} or extending and exploiting the initial result \cite{Fitzpatrick:2014vua,Alday:2015eya,Kaviraj:2015cxa,Li:2015itl,Komargodski:2016gci,Alday:2016njk,Simmons-Duffin:2016wlq,Kusuki:2021gpt,Numasawa:2022cni,Pal:2023cgk}. In particular, the Lorentzian inversion formula for four-point correlators encoded the initial observation of \cite{Komargodski:2012ek,Fitzpatrick:2012yx} in a general framework which allows to extract the CFT data from the discontinuity of the correlator, thus providing an efficient computational method for all those theories that admit a perturbative expansion and some control on the perturbative spectrum \cite{Caron-Huot:2017vep}.
\bigskip

The line of development we are interested in is the extension of the lightcone bootstrap program to multipoint correlators \cite{Antunes:2021kmm,Kaviraj:2022wbw,Harris:2024nmr}. Extending the bootstrap program to higher-point correlation functions provides access to a larger set of constraints and to a new set of exchanged local operators \cite{Rosenhaus:2018zqn,Parikh:2019ygo,Fortin:2019dnq,Goncalves:2019znr,Parikh:2019dvm,Pal:2020dqf,Fortin:2020bfq,Bercini:2020msp,Buric:2020dyz,Fortin:2020zxw,Poland:2021xjs,Buric:2021ywo,Buric:2021ttm,Buric:2021kgy,Fortin:2022grf,Poland:2023vpn,Fortin:2023xqq,Antunes:2023kyz,Poland:2023bny,Harris:2025cxx,Antunes:2025vvl}. In particular, the lightcone bootstrap for four identical scalar operators gives access to the large-spin spectrum of double-twist operators and their OPE coefficients with the external scalars. On the other hand, considering five- and six-point functions significantly enlarges the set of CFT data, providing access to OPE coefficients of spinning operators as well as the spectrum and OPE coefficients of triple-twist operators. The drawback is that the kinematics of higher-point functions is significantly  more intricate and this makes it harder to derive closed-form expressions for the CFT data.
\bigskip

In this paper, we analyze a setup that enlarges the set of available CFT data with a milder complication of the kinematical structure. To achieve this, we consider a CFT in the presence of a conformal defect, i.e. an extended excitation which preserves conformal invariance along its profile. Starting with the seminal work of \cite{Billo:2016cpy}, the application of bootstrap techniques to conformal defects has evolved into an independent research line, the defect bootstrap program. If one is interested in the interaction between bulk and defect degrees of freedom, the first observable with a non-trivial crossing equation is the two-point function of bulk operators. The lack of positivity in the bulk channel does not allow for a straightforward application of numerical bootstrap methods (see however \cite{Meineri:2025qwz,Lanzetta:2025xfw} for recent progress), but it is certainly possible to apply analytic bootstrap techniques. In particular, Lorentzian inversion formulae \cite{Lemos:2017vnx,Liendo:2019jpu} and dispersion relations \cite{Bianchi:2022ppi,Barrat:2022psm}\footnote{See also \cite{Bonomi:2024lky} for the case of four operators completely lying within a line defect.} have been derived for both OPE channels and their simplicity allowed for the computation of the correlator in situations where one can expand in a small parameter having some control on the spectrum of exchanged operators \cite{Bissi:2018mcq,Dey:2020jlc,Barrat:2021yvp,Liendo:2016ymz,Gimenez-Grau:2019hez,Gimenez-Grau:2021wiv,Bianchi:2022sbz,Gimenez-Grau:2022ebb,Bianchi:2023gkk,Chen:2023yvw,Chen:2024orp}. 
\bigskip

\begin{figure}[htp]
    \centering
    \includegraphics[width=\linewidth]{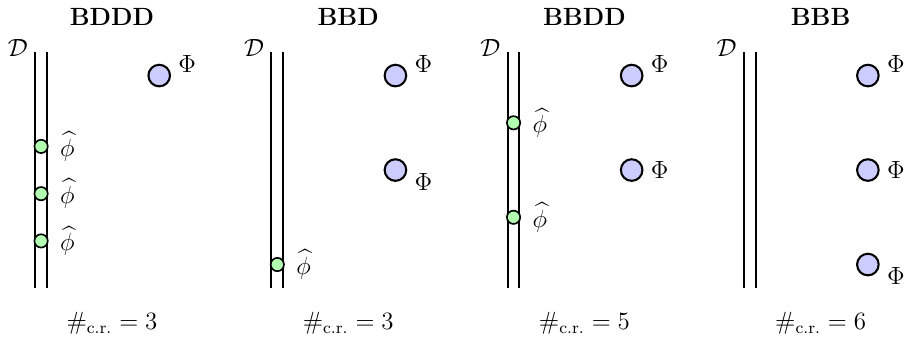}
    \caption{Examples of multipoint defect correlators with their respective number of conformal cross-ratios.}
    \label{fig:higher-point-cases}
\end{figure}

Here, we are interested in extending Lorentzian bootstrap methods to multipoint defect correlators. One of the nice features of defect correlators is that the set of available configurations is extremely rich. One option, of course, is to consider several defect operators (see, for instance, \cite{Artico:2024wut} for a perturbative application in superconformal field theories), but we would like to probe also the interaction with the bulk. Some higher-point functions with at least one bulk operator are shown in Figure~\ref{fig:higher-point-cases}, where we also indicate the number of conformal cross-ratios associated to each correlator. The number of cross-ratios could be used as a measure of the complexity of the kinematical structure. To compare with the homogeneous case, we should remember that the four-point function of local operators is a function of two cross-ratios (exactly like the two-point function of bulk operators with a defect), while the five- and six-point functions can be expressed in terms of five and nine cross-ratios, respectively. As we see in Figure \ref{fig:higher-point-cases}, for defect correlators the difficulty grows at a slower rate as we can construct intermediate observables which are just slightly more complicated than the bulk two-point function. Among those, the simplest examples are the BDDD and the BBD correlators\footnote{One may wonder why the BDD correlator is not included in this list. It turns out that the BDD correlator is even simpler than the BB one as it is characterized by a single cross-ratio and there is no crossing equation (see however \cite{Artico:2024wnt,Bartlett-Tisdall:2025iqx} for recent analyses of this setup or~\cite{Chen:2023yvw,Levine:2023ywq,Levine:2024wqn,Loparco:2025aag} for the boundary ($\partial$) cases B$\partial\partial$ and BB$\partial$).}. The former however, does not probe new bulk-to-defect CFT data and for this reason in this paper we focus on the latter. 

\subsection*{Summary of the results}
The main goal of this work is to extract information about the spectrum and OPE coefficients of defect operators with large spin by requiring consistency of the crossing equation in the lightcone limit. The class of defect operators that can be constrained using the bulk two-point function of two identical scalars $\expval{\Phi\Phi}$ are the so-called transverse derivative operators and they take the schematic form  $[\Phi]_{l,s}=\partial_{\perp}^s \square_{\perp}^l \Phi$, where the operator $\Phi$ is taken on the defect and the derivatives act on the orthogonal directions \cite{Lemos:2017vnx}. Using consistency of the crossing equation in the lightcone limit one can prove that any defect conformal field theory must contain this family of operators in their defect spectrum and their half-twist at infinite spin is $\hat{h}=\frac{\hat{\Delta}-s}{2}=\frac{\Delta_{\Phi}}{2}+m$. 

\begin{figure}[htbp]
    \centering
    \includegraphics[width=0.7\linewidth]{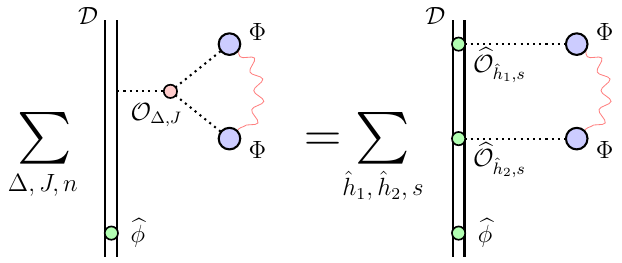}
    \caption{A schematic representation of the crossing equation for the bulk-bulk-defect correlator. The red curly line indicates the $\bar z\to 1$ lightcone limit we are considering, where the two bulk operators become lightlike separated. The defect operator $\hat{\phi}$ acts as a spectator in both OPE channel decompositions.}
    \label{fig:BBD-crossing}
\end{figure}

Here we consider the correlator $\big\langle\Phi \Phi \hat \phi\big\rangle$, where we added a defect operator $\hat{\phi}$. This correlator admits two different OPE expansions, which are schematically represented in Figure \ref{fig:BBD-crossing}. We study the resulting crossing equation in the limit where the two bulk operators become lightlike separated as shown in Figure \ref{fig:BBD-crossing} through a red curly line. In the bulk channel, the exchanged operator with lowest twist provides the leading contribution in the lightcone limit. A crucial difference compared to the BB case is that the bulk identity operator is not exchanged as it does not couple with a generic defect operator $\hat{\phi}$ (unless $\hat{\phi}$ is the defect identity of course). Therefore, to understand which operators dominate the bulk expansion at low twist we need some insight into the twist spectrum of the bulk theory. Focusing on unitary theories, we know that the unitarity bounds for the bulk theory impose $h\geq \frac{d-2}{4}$ for scalars and $h\geq \frac{d-2}{2}$ for spinning operators (remember that $h=\frac{\Delta-J}{2}$ is the half twist, so for the stress tensor $h=\frac{d-2}{2}$ and for a scalar $h=\frac{\Delta}{2}$). In this work we exclude higher-spin conserved currents as they are expected to be absent in an interacting CFT for $d>2$ \cite{Maldacena:2011jn}. Therefore, there are no spinning operators saturating the unitarity bound for $J>2$. The leading-twist contribution is given either by a conserved spinning operator such as global currents and the stress tensor, or by a light scalar operator (with $\frac{d-2}{2}\leq\Delta\leq d-2$). In this work we analyze these options and we study the implications for the defect spectrum at large spin.

In the defect channel, the exchanged operators $\hat{\mathcal{O}}_{\hat{h}_1,s}$ and $\hat{\mathcal{O}}_{\hat{h}_2,s}$ must have the same transverse spin, which is taken large to reproduce the bulk channel, but they have different transverse twists $\hat{h}_1$ and $\hat{h}_2$. 
In particular, to reproduce the lowest-twist contribution in the bulk channel we find that the defect channel must contain, at large transverse spin, two families of pairs of exchanged operators, both built out of operators that are already present in the spectrum of a generalized free field (GFF) in the presence of the defect. The first family consists of two transverse-derivative operators\footnote{With a slight abuse of language, we will often refer to the transverse-derivative operators, with $\hat{h}=\frac{\Delta_\Phi}{2}+l$, as ``single-twist'' operators, and to those with $\hat{h}=\frac{\Delta_\Phi}{2}+\frac{\hat{\Delta}_{\hat\phi}}{2}+q+l$ as ``double-twist'' operators.}
\begin{equation}
    [\Phi]_{l,s}=\Box_{\perp}^{\,l}\,\partial_{\perp}^{\,s}\,\Phi\,,\qquad \hat{h}=\frac{\Delta_{\Phi}}{2}+l\,.
\end{equation}
The second family consists of  the exchange of a transverse-derivative operator $[\Phi]_{l,s}$ together with a ``double-twist'' operator of the schematic form
\begin{equation}
    [\hat{\phi}\Phi]_{q,l,s}=\hat{\phi}\,\Box_{\parallel}^{\,q}\,\Box_{\perp}^{\,l}\,\partial_{\perp}^{\,s}\,\Phi\,,\qquad \hat{h}=\frac{\Delta_{\Phi}}{2}+\frac{\hat{\Delta}_{\hat{\phi}}}{2}+q+l\,,
\end{equation}
which involves the same transverse-derivative operator, but includes additional $q$ boxes parallel to the defect. In GFF, the three-point couplings of two transverse-derivative operators with $\hat\phi$ vanish, and the double-twist operators do not couple to the bulk field $\Phi$. The two families therefore encode the first corrections to GFF defect data: the three-point couplings $\hat{\lambda}_{[\Phi]_{l+m,s}[\Phi]_{l,s}\hat\phi}$, where the hat distinguishes OPE coefficients of defect operators from bulk ones, and the bulk-to-defect couplings $b_{\Phi[\hat{\phi}\Phi]_{q,l,s}}$.

Requiring consistency of the crossing equation in the lightcone limit we can then determine the large-spin behavior of these CFT data in terms of the product $\lambda_{\Phi \Phi \mathcal{O}_{\star}}b_{\mathcal{O}_{\star}\hat{\phi}}$ of a bulk OPE coefficient and a bulk-to-defect coupling of the leading-twist operator $\mathcal{O}_\star$. The crossing equation is only sensitive to the products $b_{\Phi \widehat{\mathcal{O}}_1}b_{\Phi \widehat{\mathcal{O}}_2} \hat{\lambda}_{\widehat{\mathcal{O}}_1\widehat{\mathcal{O}}_2\hat{\phi}}$, but in each family the remaining factors are GFF data, known from the bulk two-point function $\langle\Phi\Phi\rangle$ and from the defect four-point function $\langle\hat\phi[\Phi]\hat\phi[\Phi]\rangle$, so that the new couplings can be isolated. For a scalar bulk exchange of dimension $\Delta_\star=2h_\star$ and, for simplicity, no transverse boxes on the second operator, $l=0$, we find
\begin{align}
    \hat{\lambda}_{[\Phi]_{m,s}[\Phi]_{0,s}\hat\phi}&\stackrel{s\to\infty}{\simeq}\lambda_{\Phi\Phi\mathcal{O}_\star}\, b_{\mathcal{O}_\star\hat\phi}\,\frac{\Gamma(2h_\Phi)\Gamma(2h_\star)}{\Gamma(2h_\Phi-h_\star)\Gamma(h_\star)^{2}}\binom{m-\frac{d}{2}+2h_\Phi}{m}^{-\frac12}\notag\\
    &\hspace{130pt}\times\frac{(1-h_\star)_{m}}{m!}\,\frac{\Gamma(\hat{h}_{\hat{\phi}}-m)\Gamma(\hat{h}_{\hat{\phi}}+m)}{\Gamma(2\hat{h}_{\hat{\phi}})}\;s^{-h_\star}\,,\label{eq:intro-lambda-family-I}\\
    b_{\Phi[\hat{\phi}\Phi]_{q,0,s}}&\stackrel{s\to\infty}{\simeq}\lambda_{\Phi\Phi\mathcal{O}_\star}\, b_{\mathcal{O}_\star\hat\phi}\,\frac{\Gamma(2h_\Phi)^{\frac12}\Gamma(2h_\star)\Gamma(-\hat{h}_{\hat{\phi}})}{\Gamma(2h_\Phi-h_\star)\Gamma(h_\star)\Gamma(h_\star-\hat{h}_{\hat{\phi}})}\notag\\
    &\hspace{60pt}\times\,\frac{(-1)^{q}(1-h_\star+\hat{h}_{\hat{\phi}})_{q}}{(1+\hat{h}_{\hat{\phi}})_{q}}\left[\frac{(2\hat{h}_{\hat{\phi}})_{q}\,(\frac{p}{2})_{q}}{q!\,(2\hat{h}_{\hat{\phi}}-\frac{p}{2}+1)_{q}}\right]^{\frac12} s^{h_\Phi-h_\star-\frac12}\,,\label{eq:intro-b-family-II}
\end{align}
where $h_\Phi=\frac{\Delta_\Phi}{2}$, $\hat{h}_{\hat\phi}=\frac{\hat\Delta_{\hat\phi}}{2}$, $h_\star$ is the half twist of $\mathcal{O}_\star$ and $p$ is the dimension of the defect. Both sets of couplings are suppressed by a factor $s^{-h_\star}$ with respect to the corresponding GFF data, $b^{\text{GFF}}_{\Phi[\Phi]_{l,s}}\sim s^{h_\Phi-\frac12}$, with a falloff dictated by the twist of the leading bulk exchange. No other pair of defect operators is required at this order: in particular, no operators with twist higher than the double-twist ones appear. The expressions for generic $l$ and for spinning leading-twist exchanges are given in~\eqref{eq:lambda-family-I} and~\eqref{eq:b-family-II}.

These results are very general and they hold for any unitary interacting CFT. As an application, we analyzed the specific case of the BBD three-point function of two chiral primary operators in the bulk and one tilt operator on the defect in $\mathcal{N}=4$ SYM. In that case, the spectrum of exchanged operators in the bulk channel is well known and the leading-twist contribution comes from operators in the stress tensor supermultiplet. When specified to that case, our results significantly simplify and we collect them in Section \ref{section:Applications}.

This paper is structured as follows. Section~\ref{section:Defect_CFT} reviews the standard lightcone bootstrap techniques for the BB two-point function. The main results of our work mentioned in this introduction are presented in Sections~\ref{section:Multipoint_dCFT} and~\ref{section:Applications}. In Section~\ref{section:Outlook} we outline future directions. Finally, a number of technical details are collected in Appendices~\ref{section:embedding}--\ref{sec:solving-crossing}. 

\paragraph{Note added.} In the published version of this paper the two parameters $c_1$ and $c_2$ of the Appell $F_4$ function in the defect-channel block~\eqref{eq:appell F4} were interchanged. This affected the large-spin defect blocks and, consequently, the solution of the crossing equation in Section~\ref{section:Multipoint_dCFT}, the results of Section~\ref{section:Applications} and the summary in the Introduction, which have all been corrected in the present version.

\section{Review: bulk two-point correlators and lightcone bootstrap}
\label{section:Defect_CFT}
\begin{figure}[htp]
    \centering
    \includegraphics[width=0.5\linewidth]{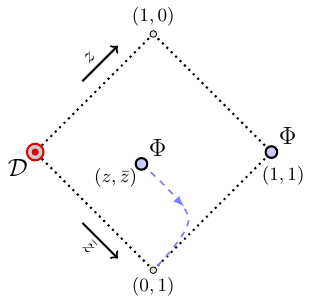}
    \caption{The role of the $z,\bar z$ cross ratios can be easily understood in the conformal frame where the two bulk operators lie on a Minkowski plane orthogonal to the defect $\mathcal{D}$. The $z=0$ and $\bar z=0$ lines represent the defect lightcone, while the $z=1$ and $\bar z=1$ lines correspond to the two bulk operators being lightlike separated.}
    \label{fig:2ptconframe}
\end{figure}

To set the stage for higher-point applications, let us review how the lightcone bootstrap works for bulk two-point functions \cite{Lemos:2017vnx}. We consider a flat $p$-dimensional defect, and we denote the codimension of the defect with $q=d-p$. We will always assume that $q>1$ \footnote{For $q=1$ the kinematics is different and there is no notion of transverse spin.}. We work in the embedding space formalism, which we review in Appendix \ref{section:embedding}. The bulk two-point function of identical scalar operators reads
\begin{equation}
 \expval{\Phi_{\Delta_\Phi}(P_1)\Phi_{\Delta_\Phi}(P_2)}=\frac{f(\xi, \cos\varphi)}{\left(P_1\trasc P_1\right)^{\frac{\Delta_\Phi}{2}}\left(P_2\trasc P_2\right)^{\frac{\Delta_\Phi}{2}}}
 \label{eq:2bulk correlator}
 \end{equation}
with the cross ratios
\begin{equation}
    \xi = - \frac{2 P_1 \cdot P_2}{\left(P_1\trasc P_1\right)^{\frac12}\left(P_2\trasc P_2\right)^{\frac12}}\,, \qquad \cos \varphi = \frac{P_1\trasc P_2}{\left(P_1\trasc P_1\right)^{\frac12}\left(P_2\trasc P_2\right)^{\frac12}}\,.
\end{equation}
To understand the geometry of the problem in Lorentzian signature it is convenient to consider the conformal frame depicted in Figure \ref{fig:2ptconframe}.
The convenient cross ratios in that frame are $z$ and $\bar z$ given by
\begin{equation}
    \xi=\frac{(1-z)(1-\bar z)}{(z\bar z)^{\frac12}}\,, \qquad \cos \varphi=\frac{z+\bar z}{2 (z\bar z)^{\frac12}}\,.
    \label{eq:zbarz_def}
\end{equation}

\begin{figure}[htp]
    \centering
    \includegraphics[width=0.7\linewidth]{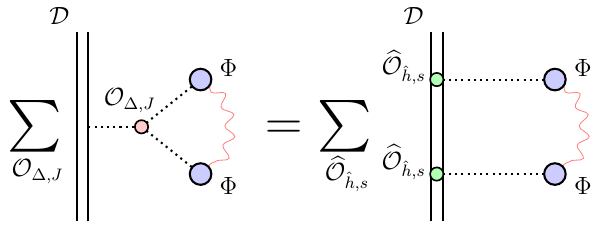}
    \caption{A schematic representation of the crossing equation for the bulk two-point function. The red curly line indicates the $\bar z\to 1$ lightcone limit we are considering, where the two bulk operators become lightlike separated.}
    \label{fig:BB-crossing}
\end{figure}

The function $f(z,\bar z)$ admits two different block expansions as shown in Figure~\ref{fig:BB-crossing}.
In the defect channel, we can express the correlator as
\begin{equation}\label{eq:defectexp}
    f(z,\bar z)=\sum_{\hat{h},s} b^2_{\hat{h},s} \hat{f}_{\hat{h},s}(z,\bar z)\,,
\end{equation}
where the defect conformal blocks $\hat{f}_{\hat{h},s}(z,\bar z)$ are known explicitly \cite{Billo:2016cpy,Lemos:2017vnx}
\begin{equation}
    \hat{f}_{\hat{h},s}(z,\bar{z})=z^{\hat{h}}\bar{z}^{\hat{h}+s}{}_{2}F_{1}\left( -s,\frac{q}{2}-1,2-\frac{q}{2}-s,\frac{z}{\bar{z}}\right)\,{}_{2}F_{1}\left( 2\hat{h}+s,\frac{p}{2},2\hat{h}+s+1-\frac{p}{2},z\bar{z}\right)\,.
    \label{eq:defectblock}
\end{equation}
In \eqref{eq:defectexp} we are summing over the transverse spin $s$ and the half transverse twist $\hat{h}=\frac{\hat{\Delta}-s}{2}$ of the defect exchanged operator. The coefficients $b_{\hat{h},s}$ are bulk-to-defect couplings determined by
\begin{equation}
    \expval{\Phi_{\Delta_{\Phi}}(P_1) \widehat{\mathcal{O}}_{\hat{h},s}(P_2,W)}=\frac{b_{\hat{h},s}(P_1\trasc W)^s}{(P_1\parsc P_2)^{2\hat{h}+s} (P_1\trasc P_1)^{\frac{\Delta_{\Phi}}{2}-\hat{h}}}\,.
\end{equation}

In the bulk channel, the correlator can be expanded as
\begin{equation} \label{eq:bulkexp}
    f(z,\bar z)=\left(\frac{(1-z)(1-\bar z)}{(z\bar z)^{\frac12}}\right)^{-{\Delta}_{\Phi}}\sum_{\Delta,J} \lambda_{\Phi\Phi \mathcal{O}} \,a_{\mathcal{O}} f_{\Delta,J}(z,\bar z)\,,
\end{equation}
where we sum over the dimension $\Delta$ and the spin $J$ of the bulk exchanged operator. The bulk conformal blocks $f_{\Delta,J}(z,\bar z)$ can be expressed as solutions to a Casimir differential equation as we review in Appendix \ref{section:bulk-channel-blocks}. The coefficients $\lambda_{\Phi\Phi \mathcal{O}}$ are bulk three-point functions, while the coefficients $a_{\mathcal{O}}$ appear in the one-point function 
\begin{equation}
    \expval{\mathcal{O}_{\Delta,J}(P,Z)}=a_\mathcal{O}\frac{\left(\frac{(P\circ Z)^2}{P\circ P}-Z\trasc Z\right)^{\frac{J}{2}}}{\left(P\trasc P\right)^{\Delta+J}}\,.
\end{equation}

We are interested in a regime where 
\begin{equation}
    1-\bar z\ll z <1\,,
\end{equation}
which corresponds to a double lightcone limit approached in a specific direction as depicted in Figure \ref{fig:2ptconframe}. Specifically, we will work at leading order for $\bar z \to 1$ and at all orders for $z\to 0$. At this point, it is important to mention that the procedure we follow in this paper, both for this review and for the higher-point generalization, is not completely rigorous as it was addressed in a series of papers \cite{Qiao:2017xif,Kravchuk:2020scc,Kravchuk:2021kwe,Pal:2022vqc,vanRees:2024xkb} for the homogeneous case. In this first work on the subject, we neglect these subtleties and we assume that all the nice features of the OPE expansion in the double lightcone limit that worked for the homogeneous case keep working also for the defect case. 

In the limit $\bar z \to 1$ the bulk expansion converges and we can use \eqref{eq:bulkexp} to evaluate $f(z,\bar z)$ for $\bar z\to 1$. In particular, the bulk block in the lightcone limit reads
\begin{equation}\label{eq:bulkbulklightcone}
    f_{\Delta,J}(z,\bar z)=(1-\bar z)^{\frac{\Delta-J}{2}}  \left(\tilde f_{\Delta+J}(z)+O(1-\bar z)\right)\,,
\end{equation}
where the function $\tilde f_{\Delta+J}(z)$ is given in \eqref{eq:bulkbulkblocklightcone}. This implies that the leading contribution is determined by exchanged operators with the lowest twist. When considering two identical external operators the lowest-twist exchanged operator is clearly the identity. In that case $\Delta=J=0$ and the correlator is simply given by the prefactor in \eqref{eq:bulkexp}. The crossing equation states that this contribution must be reproduced in the defect channel and this imposes constraints on the large-spin spectrum of defect operators. In particular, one can reproduce the $(1-\bar z)^{-\Delta_{\Phi}}$ factor by considering a scaling limit such that $s\to \infty$ and $\bar z \to 1$ with $s(1-\bar z)$ kept fixed. Expanding the defect blocks in this limit (see Appendix \ref{section:largespinexpansion} for details), we get
\begin{equation} \label{eq:defectblocksbulkbulklarges}
    \hat{f}_{\hat{h},s}(z,\bar z)\sim z^{\hat{h}}(1-z)^{1-\frac{d}{2}}e^{-s (1-\bar{z})}\,.
\end{equation}
To reproduce the contribution of the bulk identity order by order in $z\to 0$ we need to ask that the large-spin spectrum of the theory contains an infinite tower of operators with transverse twist 
\begin{equation}\label{eq:largespinspectrumbulkbulk}
    \hat{h}= \frac{\Delta_\Phi}{2} + m + O(s^{-\alpha})\,, \qquad s\to\infty\,,
\end{equation}
for $m\in \mathbb{N}$ and $\alpha >0$. As discussed in \cite{Lemos:2017vnx}, these operators are single-twist defect operators of the schematic form $\partial_{\perp}^s \square_{\perp}^m \Phi$. They appear naturally in GFF and their leading role at large spin is the defect counterpart of the emergence of GFF in the large-spin spectrum of homogeneous CFTs \cite{Fitzpatrick:2012yx,Komargodski:2012ek}. To reproduce the prefactor in \eqref{eq:bulkexp}, one also needs to impose that the defect OPE coefficients of these single twist operators $b^2_{m,s}$ behave as a power law for $s\to \infty$
\begin{equation} \label{eq:opecoeffbulkbulk}
b^2_{m,s}=s^{\Delta_{\Phi}-1}\left(\frac{1}{\Gamma(\Delta_{\Phi})} \left(\begin{array}{c}
        m-\frac{d}{2}+\Delta_{\Phi}   \\
        m 
    \end{array}\right) + O(s^{-\beta})\right)\,,
\end{equation}
for $\beta >0$. Inserting \eqref{eq:opecoeffbulkbulk} and \eqref{eq:defectblocksbulkbulklarges} into the defect block expansion \eqref{eq:defectexp}, we can replace the sum over spins with an integral dominated by the large-spin region such that
\begin{equation}
    f(z,\bar z) \sim \sum_{m=0}^{\infty}\left(\begin{array}{c}
        m-\frac{d}{2}+\Delta_{\Phi}   \\
        m 
    \end{array}\right) \frac{z^{\frac{\Delta_{\Phi}}{2}+m} (1-z)^{1-\frac{d}{2}}}{\Gamma(\Delta_{\Phi})}\int_{\Lambda}^\infty e^{-s(1-\bar z)} s^{\Delta_{\Phi}-1} ds\,,
\end{equation}
for some cut-off $\Lambda$, whose value is unimportant in the $\bar z \to 1$ limit. The integral, at leading order for $\bar z\to 1$, reproduces the $(1-\bar z)^{-\Delta_{\Phi}}$ factor in \eqref{eq:bulkexp} yielding
\begin{equation} \label{eq:bulbulkfinalexp}
    f(z,\bar z)\sim (1-\bar z)^{-\Delta_{\Phi}} (1-z)^{-\Delta_{\Phi}} z^{\frac{\Delta_{\Phi}}{2}}\,. 
\end{equation}
This result agrees with the prefactor in \eqref{eq:bulkexp} for $\bar z \to 1$. In principle, one could use the same strategy to reproduce the twist-subleading contributions in the bulk expansion \eqref{eq:bulkexp}. This constrains the large-spin corrections to the classical dimension \eqref{eq:largespinspectrumbulkbulk} and to the OPE coefficients \eqref{eq:opecoeffbulkbulk} and a systematic large-spin expansion can be established. We refer to \cite{Lemos:2017vnx} for further details on this expansion, which can be understood more rigorously in the context of the Lorentzian inversion formula. For the the higher-point generalization, which presents a richer and more complicated kinematical structure, we will concentrate on the leading-twist contribution. As we will highlight in the next Section, the leading contribution is already highly non-trivial because the bulk-bulk-defect correlator does not include the identity in the bulk channel.

\section{Multipoint defect correlators}
\label{section:Multipoint_dCFT}
In the previous section, we reviewed how the crossing equation for the bulk two-point correlator in a lightcone limit determines the asymptotics of bulk-to-defect OPE coefficients~$b^2_{m,s}$ at large spin. Our aim now is to study the simplest extension of these constraints in the setting of higher-point correlation functions. 
The scalar higher-point correlators with simplest kinematics were summarized in Figure~\ref{fig:higher-point-cases}, where we also represented the number of cross-ratios associated with each case. 
As we argued there, the simplest non-trivial case which admits a crossing equation involving both bulk expansions and defect expansions is the Bulk-Bulk-Defect (BBD) correlator, which will be the main focus of the rest of this work.

\subsection{The Bulk-Bulk-Defect bootstrap setup}
\label{sec:BBD-setup}
\begin{figure}[htp]
    \centering
    \includegraphics[width=0.5\linewidth]{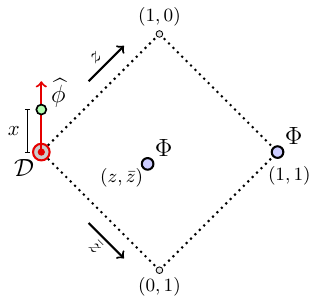}
    \caption{The role of the $z,\bar z$ cross ratios is the same as for the case of the bulk two-point function. The two bulk operators lie on a Minkowski plane orthogonal to the defect $\mathcal{D}$. The defect operator $\hat{\phi}$ is located along the defect at a distance $x$ from the transverse plane.}
    \label{fig:BBD-frame-nolimit}
\end{figure}

A BBD correlator can be expressed as follows
\begin{equation}
    \expval{\Phi(P_1)\Phi(P_2)\hat\phi(P_3)}=\Omega_{\Delta_{\Phi}\hat{\Delta}_{\hat\phi}}\!(P_1,P_2,P_3)\, \mathcal{G}(z,\bar{z},x)\,,
\end{equation}
in terms of a conventional prefactor,
\begin{equation}
    \Omega_{\Delta_{\Phi}\hat{\Delta}_{\hat\phi}}=\frac{(-2P_{1}\bullet P_{2})^{\hat{\Delta}_{\hat\phi}/2}}{(-2P_{1}\bullet P_{3})^{\hat{\Delta}_{\hat\phi}/2}(-2P_{2}\bullet P_{3})^{\hat{\Delta}_{\hat\phi}/2}(P_{1}\circ P_{1})^{\Delta_{\Phi}/2}(P_{2}\circ P_{2})^{\Delta_{\Phi}/2}}\,,
    \label{eq:Omega_prefactor}
\end{equation}
and a function of three cross-ratios $\mathcal{G}(z,\bar{z},x)$. Two of these cross-ratios, $z$ and $\bar{z}$, are identical to the Bulk-Bulk cross-ratios defined in~\eqref{eq:zbarz_def}, while the third one, $x$, can be defined via
\begin{equation}
    \frac{x^2+z\bar{z}}{(1+x^2)(1-z)(1-\bar{z})}=-\frac{\left(P_1\trasc P_1\right)\left(P_2\parsc P_3\right)}{2\left(P_1\cdot P_2\right)\left(P_1\parsc P_3\right)}\,.
\end{equation}
These three degrees of freedom can be easily visualized in the conformal frame of Figure~\ref{fig:BBD-frame-nolimit}
where $x$ is the distance of the defect field insertion $\hat\phi$ from the $(z,\bar z)$-plane where the two bulk fields $\Phi$ are located.

As before, we can expand the correlator in two different OPE channels as represented in Figure~\ref{fig:BBD-crossing}.
The bulk-channel expansion has the form
\begin{equation}
    \mathcal{G}(z,\bar z,x)=\sum_{\Delta,J,n}\lambda_{\Phi\Phi\mathcal{O}} \,b_{\mathcal{O}\hat\phi}^{(n)}\,f_{\Delta,J,n}\,(z,\bar z,x)
\end{equation}
where $\Delta\in \mathds{R}_+$ and $J\in \mathbb{N}$ are the labels of the exchanged bulk primaries in the $\Phi\times\Phi$ OPE, while $n=0,\dots,\lfloor J/2\rfloor$ labels the possible tensor structures of the bulk-to-defect two-point function\footnote{We remind the reader that our embedding space conventions, including the definition of the $C_i^{AI}$, are gathered in Appendix~\ref{section:embedding}.}
\begin{equation}
    \expval{\!\mathcal{O}_{\Delta,J}(P_1,Z_1) \hat{\phi}(P_2)\!}=\frac{\left(4 C_1^{AI}P^2_A  P^1_I\right)^{J}}{\left(-2 P_1\parsc P_2\right)^{\hat\Delta_{\hat\phi}+J}\!\left(P_1\trasc P_1\right)^{\frac{\Delta-\hat\Delta_{\hat\phi} +J}{2}}} \sum_n b_{\mathcal{O}\hat\phi}^{(n)}\!\left(\!\frac{C_1^{AI} C^1_{AI}\!\left(P_1\parsc P_2\right)^2}{2\left(C_1^{AI}P^2_A  P^1_I\right)^2}\!\right)^{\lfloor\frac{J}{2}\rfloor-n}\hspace*{-5pt}.
    \label{eq:Bulk-to-defect_with_bulk_spin}
\end{equation}
Note that here we are considering general contributions for a generic $p$-dimensional defect. We are therefore not considering special cases where parity-odd tensor structures can contribute, for which the expressions in~\eqref{eq:Bulk-to-defect_with_bulk_spin} would only constitute a part of the answer.

The defect channel, instead, can be expressed as
\begin{equation}
    \mathcal{G}(z,\bar z,x)=\sum_{\hat{h}_1,\hat{h}_2,s}b_{\Phi\widehat{\mathcal{O}}_{1}}b_{\Phi\widehat{\mathcal{O}}_{2}}\hat{\lambda}_{\widehat{\mathcal{O}}_{1}\widehat{\mathcal{O}}_{2}\hat{\phi}}\, \hat{f}_{\hat{h}_1,\hat{h}_2,s}(z,\bar z,x)\,.
\end{equation}
Here, the labels are the half transverse twists $\hat{h}_1$, $\hat{h}_2$ and the transverse spin $s_1=s_2\equiv s$ of the two operators $\widehat{\mathcal{O}}_1$, $\widehat{\mathcal{O}}_2$, while $\hat{\lambda}_{\widehat{\mathcal{O}}_{1}\widehat{\mathcal{O}}_{2}\hat{\phi}}$ is the OPE coefficient of three defect operators, which we distinguish with a hat from the bulk OPE coefficients $\lambda_{\Phi\Phi\mathcal{O}}$.

We aim to study the crossing equation
\begin{equation}
    \sum_{\Delta,J,n}\lambda_{\Phi\Phi\mathcal{O}} \,b_{\mathcal{O}\hat\phi}^{(n)}\,f_{\Delta,J,n}(z,\bar z,x)=\sum_{\hat{h}_1,\hat{h}_2,s}b_{\Phi\widehat{\mathcal{O}}_{1}}b_{\Phi\widehat{\mathcal{O}}_{2}}\hat{\lambda}_{\widehat{\mathcal{O}}_{1}\widehat{\mathcal{O}}_{2}\hat{\phi}}\, \hat{f}_{\hat{h}_1,\hat{h}_2,s}(z,\bar z,x)
\end{equation}
in the limit $\bar{z}\to1$, where the two bulk fields become null separated. This is a natural limit to consider in the bulk channel as it corresponds to the Lorentzian OPE limit, in which the expansion is dominated by the operators with lowest twist $2h=\Delta-J$.

\paragraph{The leading-twist exchange} Contrary to the bulk-bulk case, the identity exchange is not allowed in the bulk channel of the BBD correlator because the one-point function $\langle\hat{\phi}(P_3)\rangle$ is zero (or, analogously, because $b_{\mathds{1}\hat{\phi}}=0$). This means that the first contribution in the $\bar{z}\to 1$ lightcone limit is theory-dependent and thus already non-trivial. 
This leading contribution corresponds to the allowed exchanged operators with the lowest possible twist above the identity $h_\star>0$. We will work under the assumption that there is only one such operator with lowest twist $h_\star$, which we dub the leading-twist operator $\mathcal{O}_\star$. This assumption is not very restrictive, as the case of a finite number of leading-twist operators with the same twist is simply covered by running our analysis term-by-term and recombining all contributions in the end. 
We will be considering the spin $J$ of the leading-twist operator to be even, as this is the only case compatible with Bose symmetry of the external bulk fields\footnote{\label{footnote_antisymmetric} If the fields are in non-trivial representations of a global symmetry, the constraint from Bose symmetry relaxes and odd-spin exchanges are also allowed, with an antisymmetric symmetry structure. In these spin-odd cases, our subsequent formulas still apply, modulo the simple replacement $2n\to 2n+1$ and the fact that the bulk channel becomes antisymmetric, rather than symmetric, under the exchange of the two bulk operators.}. 
Due to the presence of an infinite family of operators approaching the spectrum of double-twist operators $[\Phi\Phi]_{0,J}$  at large spin $J$, our assumption automatically imposes the bound $h_\star<2h_\Phi$.  
In an interacting unitary theory, the leading-twist operator typically corresponds to the stress tensor $T^{\mu\nu}$ with $2h_T=d-2$ and $J_T=2$, to a conserved current $\mathcal{J}^\mu$ with $2h_{\mathcal{J}}=d-2$ and $J_{\mathcal{J}}=1$, or to some scalar operator $\varphi$ with $\frac{d-2}{4}\le h_{\varphi}\le \frac{d-2}{2}$. Our goal is then to explore what constraints the presence of such a leading-twist operator implies in the defect channel expansion.

\paragraph{The bulk-channel expansion}
To work out the bulk-channel block $f_{\Delta,J,n}(z,\bar z,x)$ in the $\bar{z}\to 1$ lightcone limit, we use the integral expression for the leading term in the Lorentzian OPE~\cite{Ferrara:1971zy,Ferrara:1971tq,Ferrara:1972cq}
\begin{equation}
    \Phi_1\!(P_1)\Phi_2(P_2) \! \stackrel{P_{12}\rightarrow 0}{\simeq} \sum_{\mathcal{O}}  \!\frac{(-2P_{12})^{h_{\mathcal{O}}-h_1-h_2}\lambda_{\Phi_1\Phi_2\mathcal{O}}}{\mathrm{B}(\bar h_{\mathcal{O}}+h_{12},\bar h_{\mathcal{O}}+h_{21})} \!\int_{0}^1\!\! \frac{\dd t}{t^{1-\bar h_{\mathcal{O}}+h_{12}}} (1-t)^{h_{12}+\bar{h}_{\mathcal{O}}-1} \mathcal{O}(\tilde{P},\tilde{Z})+\dots
    \label{eq:lightcone-OPE}
\end{equation}
with $\tilde{P}=P_2-t(P_2-P_1)$, $\tilde{Z}=P_2-P_1$, with $\mathrm{B}$ representing the Euler beta function, and where we dropped subleading terms in $P_{12}$. Assuming the presence of a single leading-twist operator $\mathcal{O}_\star$, the sum over $\mathcal{O}$ collapses to just one term. 
Plugging this expansion in the $\expval{\Phi(P_1)\Phi(P_2)\hat\phi(P_3)}$ correlator leads to an integral of bulk-to-defect correlators~\eqref{eq:Bulk-to-defect_with_bulk_spin}, which can be evaluated in terms of a Lauricella $F_D$ function of three variables (see Appendix~\ref{section:bulk-channel-blocks} for a more detailed derivation, including integral and series representations for this function). 
After extracting the prefactor~\eqref{eq:Omega_prefactor} from this expression, we get an analytic expression for the bulk-channel expansion in the $\bar{z}\to 1$ lightcone limit
\begin{multline}
    \mathcal{G}(z,\bar{z},x)\stackrel{\bar{z}\to 1}{\simeq}\\
    \sum_{n}\lambda_{\Phi\Phi\mathcal{O}_\star}\, b^{(n)}_{\mathcal{O}_\star\hat\phi}(1-\bar z)^{h_\star-2h_\Phi}(1-z)^{\bar{h}_\star-2h_\Phi}z^{-\bar{h}_\star+h_\Phi+\hat{h}_{\hat{\phi}}}(x^2-z)^{2n}(x^2+z)^{-2n-\hat{h}_{\hat{\phi}}}\\
    \times\left(\frac{1+x^2}{1+z}\right)^{\hat{h}_{\hat{\phi}}} F_D\!\left(\bar{h}_\star,-2n,\bar{h}_\star-\hat{h}_{\hat{\phi}},2(\hat{h}_{\hat{\phi}}+n),2\bar{h}_\star;\frac{z-1}{z-x^2},\frac{z-1}{z},\frac{z-1}{z+x^2}\right)\,,
    \label{eq:bulk-channel-leading}
\end{multline}
where $\bar{h}_\star=\frac{\Delta_\star+J_\star}{2}$ is the natural label for the restriction of the conformal group on the $P_{12}=0$ null line.

Note that, because of its second argument being a negative integer, the apparent three-variable Lauricella $F_D$ function considered above is actually a finite sum of two-variable Appell~$F_1$ functions, which we define in Appendix~\ref{section:Hypergeometrics}. For instance, if we consider a scalar leading-twist operator, we have $\bar{h}_{\star}=h_\star$ and $n=0$, which lead to
\begin{multline}
    \mathcal{G}(z,\bar{z},x)\stackrel{\bar{z}\to 1}{\simeq}
    \lambda_{\Phi\Phi\mathcal{O}_\star}\, b_{\mathcal{O}_\star\hat\phi}(1-\bar z)^{h_\star-2h_\Phi}(1-z)^{h_\star-2h_\Phi}z^{-h_\star+h_\Phi+\hat{h}_{\hat{\phi}}}\\
    \times\left(\frac{1+x^2}{(x^2+z)(1+z)}\right)^{\hat{h}_{\hat{\phi}}} F_1\!\left(h_\star,h_\star-\hat{h}_{\hat{\phi}},2\hat{h}_{\hat{\phi}},2h_\star;\frac{z-1}{z},\frac{z-1}{z+x^2}\right)\,,
\end{multline}
where the dependence on just two variables is manifest.

\paragraph{The defect-channel expansion} Having expressed the leading term in the bulk-channel expansion, our next task is to understand what objects we need to sum over in the defect channel side to reproduce this term. We use the Casimir singularity trick of~\cite{Simmons-Duffin:2016wlq} to simplify this task, leveraging the fact that the defect channel conformal blocks are eigenfunctions of Casimir operators of the parallel and transverse symmetry algebras. We have
\begin{equation}
    \frac{\mathcal{C}_{\parallel,i}^2 \left(\Omega_{\Delta_{\Phi}\hat{\Delta}_{\hat\phi}}\hat{f}_{\hat{h}_1,\hat{h}_2,s}\right)}{\Omega_{\Delta_{\Phi}\hat{\Delta}_{\hat\phi}}\hat{f}_{\hat{h}_1,\hat{h}_2,s}}=(2\hat{h}_i+s)(2\hat{h}_i+s-p)\,, \qquad \frac{\mathcal{C}_{\perp,i}^2 \left(\Omega_{\Delta_{\Phi}\hat{\Delta}_{\hat\phi}}\hat{f}_{\hat{h}_1,\hat{h}_2,s}\right)}{\Omega_{\Delta_{\Phi}\hat{\Delta}_{\hat\phi}}\hat{f}_{\hat{h}_1,\hat{h}_2,s}}=s(s+d-p-2)\,,
    \label{eq:defect-Casimir-eigenvalues}
\end{equation}
with $i=1,2$, where 
\begin{equation}
    \mathcal{C}_{\parallel,i}^2=\frac12 \mathcal{T}_{AB}^{(i)}\mathcal{T}^{BA}_{(i)}\,, \qquad \mathcal{C}_{\perp,i}^2=\frac12 \mathcal{T}_{IJ}^{(i)}\mathcal{T}^{IJ}_{(i)}\,,
\end{equation}
while $\mathcal{T}^{(i)}_{\mu\nu}=(P_{i})_{\mu}\partial_{P^\nu_i}-(P_{i})_{\nu}\partial_{P^\mu_i}$ is the differential action of the conformal generators on scalar fields.

We can now act with the Casimir operators on the full correlator in either the bulk channel or the defect channel expansions. From the action on the bulk channel~\eqref{eq:bulk-channel-leading} we can deduce
\begin{equation}
    \frac{\mathcal{C}_{\parallel,i}^2 \left(\Omega_{\Delta_{\Phi}\hat{\Delta}_{\hat\phi}}\mathcal{G}\right)}{\Omega_{\Delta_{\Phi}\hat{\Delta}_{\hat\phi}}\mathcal{G}}\propto (1-\bar{z})^{-2}\,, \qquad  \frac{\mathcal{C}_{\perp,i}^2 \left(\Omega_{\Delta_{\Phi}\hat{\Delta}_{\hat\phi}}\mathcal{G}\right)}{\Omega_{\Delta_{\Phi}\hat{\Delta}_{\hat\phi}}\mathcal{G}}\propto (1-\bar{z})^{-2}\,,
    \label{eq:Casimir_singularity-bulk}
\end{equation}
so the action of the Casimir operators on the correlator in the lightcone limit increases its divergent behavior for $\bar{z}\to 1$, while~\eqref{eq:defect-Casimir-eigenvalues} remains valid block by block. These Casimir operators can be applied an arbitrary number of times\footnote{Here, we are assuming generic values for the exponent of $(1-\bar{z})^{h_\star-2h_{\Phi}}$. Our arguments do not apply to the case where this exponent is a non-negative integer.} on the crossing equation, making the correlator on one side arbitrarily singular. By crossing symmetry, we can then understand that the main contribution in the defect channel expansion has to come from blocks whose transverse spin scales as $s\propto (1-\bar{z})^{-1}$, matching the divergent behavior on both sides.

From~\cite{Buric:2020zea}, the general defect-channel conformal blocks are known exactly. These can be normalized in the OPE limit and expressed in our conventions as 
\begin{equation}
    \hat{f}_{\hat{h}_1,\hat{h}_2,s}=\mathcal{N}\frac{(-v_1)^{\hat{h}_2+\frac{s}{2}}(-v_2)^{\hat{h}_1+\frac{s}{2}}}{\left(1-v_1-v_2\right)^{\hat{h}_{\hat\phi}}}F\!\left(v_{1},v_{2}\right)C_{s}^{\left(\frac{q-2}{2} \right)}(\cos{\varphi})\,,
    \label{eq:general-defect-blocks}
\end{equation}
where $\mathcal{N}=\frac{s!}{\left( \frac{q-2}{2}\right)_{s}}$ is a normalization factor, $C_{n}^{(m)}(x)$ is a Gegenbauer polynomial, the object
\begin{equation}
   F\!\left( v_{1},v_{2}\right)=F_{4}\!\left( \hat{h}_{1}\!+\!\hat{h}_{2}\!+\!s\!-\!\hat{h}_{\hat\phi},\,\hat{h}_{1}\!+\!\hat{h}_{2}\!+\!s\!-\!\hat{h}_{\hat\phi}\!+\!\tfrac{2-p}{2},\,2\hat{h}_{2}\!+\!s\!+\!\tfrac{2-p}{2},\,2\hat{h}_{1}\!+\!s\!+\!\tfrac{2-p}{2};v_{1},v_{2}\right)
    \label{eq:appell F4}
\end{equation}
is an Appell $F_4$ function, and their set of cross-ratios is related to ours via
\begin{equation}
    \begin{split}
        v_{1}=-\frac{\left( 1+x^{2}\right)^{2}z\bar{z}}{x^{2}\left( -1+z\bar{z}\right)^{2}}\,, \qquad  v_{2}=-\frac{\left( z\bar{z}+x^{2}\right)^{2}}{x^{2}\left( -1+z\bar{z}\right)^{2}}\,, \qquad \cos \varphi=\frac{z+\bar z}{2 (z\bar z)^{\frac12}}\,.
    \end{split}
\end{equation}

As we concluded from~\eqref{eq:Casimir_singularity-bulk}, in our analysis we just need the blocks at $\bar{z}\to 1$ with $s\propto (1-\bar{z})^{-1}$. We can retrieve these from~\eqref{eq:general-defect-blocks} by rescaling $\bar{z}\to 1+ \epsilon (\bar{z}-1)$ and $s\to \frac{s}{\epsilon}$, and sending $\epsilon\to0$. The outcome of this procedure, detailed in Appendix~\ref{section:largespinexpansion}, is 
\begin{equation}
    \hat{f}_{\hat{h}_{1},\hat{h}_{2},s}(z,\bar{z},x)\overunderset{s\rightarrow\infty}{\bar{z}\rightarrow1}{\simeq} e^{-s(1-\bar{z})}x^{-2(\hat{h}_{\hat\phi}-\hat{h}_{1}+\hat{h}_{2})}z^{\hat{h}_{2}}\left(\frac{(1+x^{2})(x^{2}+z)}{1+z}\right)^{\hat{h}_{\hat{\phi}}}(1-z)^{1-\frac{d}{2}}\,.
\end{equation}
For a consistency check, we verified that this expression agrees (up to normalization) with the solution of the defect-channel Casimir equations with $\bar{z}\to 1+ \epsilon (\bar{z}-1)$ and $s\to \frac{s}{\epsilon}$ in the limit $\epsilon\to 0$; in a similar fashion to the derivation of crossed-channel blocks in~\cite{Kaviraj:2022wbw,Harris:2024nmr}.
\paragraph{The crossing equation} Combining the two expansions discussed above, we can recast the crossing equation at leading order in the $\bar{z}\to 1$ lightcone limit as
\begin{equation}
    \begin{aligned}
        &\sum_{n}\lambda_{\Phi\Phi\mathcal{O}_\star}\, b^{(n)}_{\mathcal{O}_\star\hat\phi}(1-\bar z)^{h_\star-2h_\Phi}(1-z)^{\bar{h}_\star-2h_\Phi}z^{-\bar{h}_\star+h_\Phi+\hat{h}_{\hat{\phi}}}(x^2-z)^{2n}(x^2+z)^{-2n-\hat{h}_{\hat{\phi}}}\\[-8pt]
    &\hspace{50pt}\times\left(\frac{1+x^2}{1+z}\right)^{\hat{h}_{\hat{\phi}}} F_D\!\left(\bar{h}_\star,-2n,\bar{h}_\star-\hat{h}_{\hat{\phi}},2(\hat{h}_{\hat{\phi}}+n),2\bar{h}_\star;\frac{z-1}{z-x^2},\frac{z-1}{z},\frac{z-1}{z+x^2}\right)\\[6pt]
    &=
    \sum_{\hat{h}_1,\hat{h}_2,s} b_{\Phi \widehat{\mathcal{O}}_1}b_{\Phi \widehat{\mathcal{O}}_2} \hat{\lambda}_{\widehat{\mathcal{O}}_1\widehat{\mathcal{O}}_2\hat{\phi}} e^{-s(1-\bar{z})}x^{-2(\hat{h}_{\hat\phi}-\hat{h}_{1}+\hat{h}_{2})}z^{\hat{h}_{2}}\left(\frac{(1+x^{2})(x^{2}+z)}{1+z}\right)^{\hat{h}_{\hat{\phi}}}(1-z)^{1-\frac{d}{2}}\,.
    \end{aligned}
\end{equation}
Since, as we already argued before, we are focusing on the contribution of large-spin operators, we can approximate the sum over transverse spins with an integral. This allows us to match the $\bar z$ dependence of the crossing equation by fixing the transverse-spin asymptotics of the OPE coefficients:
\begin{equation}
\begin{aligned}
    (1-\bar{z})^{h_\star-2h_\Phi}=\int_0^{\infty} \dd s \, b_{\Phi \widehat{\mathcal{O}}_1}b_{\Phi \widehat{\mathcal{O}}_2}& \hat{\lambda}_{\widehat{\mathcal{O}}_1\widehat{\mathcal{O}}_2\hat{\phi}} e^{-s(1-\bar{z})}\\
    &\rotatebox[origin=c]{-45}{$\Longrightarrow$}\\[-10pt]
&\qquad b_{\Phi \widehat{\mathcal{O}}_1}b_{\Phi \widehat{\mathcal{O}}_2} \hat{\lambda}_{\widehat{\mathcal{O}}_1\widehat{\mathcal{O}}_2\hat{\phi}}\stackrel{s\to\infty}{\simeq} C_{\widehat{\mathcal{O}}_1\widehat{\mathcal{O}}_2}\frac{s^{2h_{\Phi}-h_\star-1}}{\Gamma(2 h_{\Phi}-h_\star)}\,.
\end{aligned}
\label{eq:zbar-part-solved}
\end{equation}
Implementing this and bringing all $\hat h_1$- and $\hat h_2$-independent pieces on the LHS, we are then left with a simplified crossing equation
\begin{equation}
    \begin{split}
        &\sum_{n}\lambda_{\Phi\Phi\mathcal{O}_\star}\, b^{(n)}_{\mathcal{O}_\star\hat\phi}\,(1-z)^{\bar{h}_\star-2h_\Phi+\frac{d}{2}-1}z^{-\bar{h}_\star+h_\Phi+\hat{h}_{\hat{\phi}}}\left(\frac{x^{2}}{(x^{2}+z)^{2}}\right)^{\hat{h}_{\hat{\phi}}}\left(\frac{x^{2}-z}{x^{2}+z}\right)^{2n}\\
    &\hspace{50pt}\times F_D\!\left(\bar{h}_\star,-2n,\bar{h}_\star-\hat{h}_{\hat{\phi}},2(\hat{h}_{\hat{\phi}}+n),2\bar{h}_\star;\frac{z-1}{z-x^2},\frac{z-1}{z},\frac{z-1}{z+x^2}\right)\\
    &=
    \sum_{\hat{h}_1,\hat{h}_2} C_{\widehat{\mathcal{O}}_1\widehat{\mathcal{O}}_2}\,\left(x^2\right)^{\hat{h}_{1}}\left(\frac{z}{x^2}\right)^{\hat{h}_{2}}\,,
    \end{split}
    \label{eq:crossing_simplified}
\end{equation}
in which we are yet to constrain the sum over twists $\hat{h}_1$, $\hat{h}_2$ and to determine the form of the $C_{\widehat{\mathcal{O}}_1\widehat{\mathcal{O}}_2}$ coefficients.
Due to the very simple nature of the defect-channel side of this equation, the next natural step is to express the LHS as a series expansion
in the regime $z\ll x^2 \ll 1$, i.e. first at small $z/x^2$ and then small $x^2$, corresponding to the configuration highlighted in Figure~\ref{fig:BBD-frame-limit}.
We perform this technical step explicitly in Appendix~\ref{sec:solving-crossing}, which can be skipped at a first read of this paper. In the next subsection, we finally use the crossing equation in its final form to read off the $C_{\widehat{\mathcal{O}}_1\widehat{\mathcal{O}}_2}$ coefficients by simply matching the expansion powers.

\begin{figure}[htp]
    \centering
    \includegraphics[width=0.5\linewidth]{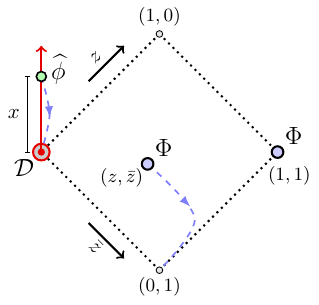}
    \caption{Here, we show the causal structure of the BBD setup in our fixed conformal frame. We highlight all the kinematic limits we consider
    in our bootstrap corresponding to the double lightcone limit $\bar{z}\to1$ and $z\to0$, together with the coplanar limit $x\to 0$, 
    in which the defect operator approaches the plane of the bulk insertions.
    The latter can also be seen as, up to the exchange of the two bulk operators, the limit in which $\Phi(P_1)$ also approaches
    the defect lightcone.}
    \label{fig:BBD-frame-limit}
\end{figure}

\subsection{Defect spectrum and CFT data at large transverse spin}
\label{sec:general-results}
As we discuss explicitly in Appendix~\ref{sec:solving-crossing}, the crossing equation~\eqref{eq:crossing_simplified} can be recast in the form
\begin{equation}
    \begin{aligned}
        \sum_{n}&\lambda_{\Phi\Phi\mathcal{O}_\star}\, b^{(n)}_{\mathcal{O}_\star\hat\phi}\,(1-z)^{\bar{h}_\star-2h_\Phi+\frac{d}{2}-1}z^{h_\Phi}\frac{\Gamma(2\bar{h}_\star)}{\Gamma(\bar{h}_\star)^{2}}\Bigg\{G^{(n)}_{0}\,{}_{2}F_{1}\!\left(\bar{h}_\star,\bar{h}_\star;1;z\right)\\
        &+\sum_{m=1}^{\infty}\frac{(1-\bar{h}_\star)_{m}}{m!}\,G^{(n)}_{m}\left[x^{2m}+\left(\frac{z}{x^{2}}\right)^{m}\right]{}_{2}F_{1}\!\left(\bar{h}_\star+m,\bar{h}_\star;1+m;z\right)\\
        &+\Gamma(\bar{h}_\star)\!\sum_{q=0}^{\infty}\frac{\Gamma(-\hat{h}_{\hat{\phi}}-q)}{\Gamma(\bar{h}_\star-\hat{h}_{\hat{\phi}}-q)}\,c^{(n)}_{q}\!\left[x^{2\hat{h}_{\hat{\phi}}+2q}\!+\!\left(\frac{z}{x^{2}}\right)^{\hat{h}_{\hat{\phi}}+q}\right]\!{}_{2}F_{1}\!\left(\bar{h}_\star\!+\!\hat{h}_{\hat{\phi}}\!+\!q,\bar{h}_\star;1\!+\!\hat{h}_{\hat{\phi}}\!+\!q;z\right)\!\Bigg\}\\
        =&\sum_{\hat{h}_1,\hat{h}_2} C_{\widehat{\mathcal{O}}_1\widehat{\mathcal{O}}_2}\,\left(x^2\right)^{\hat{h}_{1}}\left(\frac{z}{x^2}\right)^{\hat{h}_{2}}\,,
    \end{aligned}
    \label{eq:crossing-solved}
\end{equation}
where we introduced the coefficients
\begin{equation}
        c^{(n)}_{q}\equiv(-1)^{q}\sum_{j=0}^{\min(q,2n)}\binom{2n}{j}\frac{(2\hat{h}_{\hat{\phi}}+2n)_{q-j}}{(q-j)!}
        \label{eq:c-coefficients}
    \end{equation}
and
\begin{equation}
        G^{(n)}_{m}=G^{(n)}_{-m}\equiv \sum_{j=0}^{2n}(-1)^{j}\binom{2n}{j}\frac{\Gamma(\hat{h}_{\hat{\phi}}-m+j)\Gamma(\hat{h}_{\hat{\phi}}+2n+m-j)}{\Gamma(2\hat{h}_{\hat{\phi}}+2n)}\,.
        \label{eq:G-coefficients}
    \end{equation} 

Note that the contributions are organized in pairs that are symmetric under the exchange of the two bulk operators, which in the bulk lightcone limit acts on the cross-ratios as the transformation $x^2\to z/x^2$ at fixed $z$. 
By further expanding the $(1-z)$ binomial and the hypergeometric function via
\begin{equation}
    (1-z)^{\bar{h}_\star-2h_\Phi+\frac{d}{2}-1}\,{}_{2}F_{1}\!\left(\bar{h}_\star+\kappa,\bar{h}_\star;1+\kappa;z\right)=\sum_{l=0}^{\infty}P_{l}(\kappa)\,z^{l}
    \label{eq:hypergeometric_and_binomial_expansion}
\end{equation}
with
\begin{equation}
    P_{l}(\kappa)=\sum_{i=0}^{l}\frac{\left(1-\bar{h}_\star+2h_\Phi-\frac{d}{2}\right)_{i}}{i!}\,\frac{(\bar{h}_\star+\kappa)_{l-i}(\bar{h}_\star)_{l-i}}{(1+\kappa)_{l-i}\,(l-i)!}\,,
    \label{eq:P_coefficients}
\end{equation}
and then matching the powers in $x^2$ and $z/x^2$ on both sides of the crossing equation, we can identify two different families of contributions. 

\begin{itemize}
\item The contributions proportional to the $G^{(n)}_{m}$ coefficients belong to the family 
\begin{equation} 
\mathrm{(I)}\quad \Longrightarrow \quad 
    \begin{aligned}
\hat{h}_1=h_{\Phi}+l+m \quad &\Longrightarrow \quad \widehat{\mathcal{O}}_1=[\Phi]_{l+m,s}=\Box_{\perp}^{\,l+m} \partial^{i_1}_{\perp}\dots \partial^{i_s}_{\perp} \Phi\,,\\
\hat{h}_2=h_\Phi+l  \quad  &\Longrightarrow \quad \widehat{\mathcal{O}}_2=[\Phi]_{l,s}=\Box_{\perp}^{\,l}\partial^{i_1}_{\perp}\dots \partial^{i_s}_{\perp} \Phi\,,
\end{aligned}
\label{eq:exchanged_family_1}
\end{equation}
as well as its $1\leftrightarrow 2$ counterpart, which correspond to exchanges of transverse derivative operators in both legs of the defect channel. 
Performing the $z$ expansion of the binomial and hypergeometric function as in~\eqref{eq:hypergeometric_and_binomial_expansion}, the crossing equation then implies
\begin{equation}
    C_{[\Phi]_{l+m,s}[\Phi]_{l,s}}=\sum_{n}\lambda_{\Phi\Phi\mathcal{O}_\star}\, b^{(n)}_{\mathcal{O}_\star\hat\phi}\,\frac{\Gamma(2\bar{h}_\star)}{\Gamma(\bar{h}_\star)^{2}}\,\frac{(1-\bar{h}_\star)_{m}}{m!}\,G^{(n)}_{m}\,P_{l}(m)\,.
    \label{eq:C-family-I}
\end{equation}
The transverse derivative operators discussed here are already present in GFF with non-zero bulk-to-defect couplings $b_{\Phi[\Phi]_{r,s}}^{\text{GFF}}$ as determined in~\eqref{eq:opecoeffbulkbulk}, but their three-point coefficients $\hat{\lambda}_{[\Phi]_{l+m,s}[\Phi]_{l,s}\hat\phi}^{\text{GFF}}$ with a generic $\hat{\phi}$ are vanishing; this tells us that this family of contributions corresponds to the first correction to GFF where the $\hat{\lambda}_{[\Phi]_{l+m,s}[\Phi]_{l,s}\hat\phi}$ OPE coefficients are now turned on. 
We can then use the definition~\eqref{eq:zbar-part-solved} of the $C_{\widehat{\mathcal{O}}_1\widehat{\mathcal{O}}_2}$ coefficients to relate
\begin{equation}
    \hat{\lambda}_{[\Phi]_{l+m,s}[\Phi]_{l,s}\hat\phi}\stackrel{s\to\infty}{\simeq}\frac{C_{[\Phi]_{l+m,s},[\Phi]_{l,s}}}{b_{\Phi[\Phi]_{l+m,s}}\,b_{\Phi[\Phi]_{l,s}}}\,\frac{s^{2h_\Phi-h_\star-1}}{\Gamma(2h_\Phi-h_\star)}\,,
    \label{eq:lambda-family-I-general}
\end{equation}
and combining the large-spin behavior of the bulk-to-defect couplings found in~\eqref{eq:opecoeffbulkbulk} with~\eqref{eq:C-family-I}, we get
\begin{equation}
    \begin{aligned}
    \hat{\lambda}_{[\Phi]_{l+m,s}[\Phi]_{l,s}\hat\phi}\stackrel{s\to\infty}{\simeq}\frac{\Gamma(2h_\Phi)}{\Gamma(2h_\Phi-h_\star)}\,s^{-h_\star}\left[\binom{l+m-\frac{d}{2}+2h_\Phi}{l+m}\binom{l-\frac{d}{2}+2h_\Phi}{l}\right]^{-\frac12}\\
    \times \sum_{n}\lambda_{\Phi\Phi\mathcal{O}_\star}\, b^{(n)}_{\mathcal{O}_\star\hat\phi}\,\frac{\Gamma(2\bar{h}_\star)}{\Gamma(\bar{h}_\star)^{2}}\,\frac{(1-\bar{h}_\star)_{m}}{m!}\,G^{(n)}_{m}\,P_{l}(m)\,.
    \end{aligned}
    \label{eq:lambda-family-I}
\end{equation}
The three-point couplings of two transverse derivative operators with $\hat\phi$ decay therefore as $s^{-h_\star}$ at large transverse spin, with falloff dictated by the twist of the leading bulk exchange~$h_\star$, and are proportional to the product of bulk CFT data $\lambda_{\Phi\Phi\mathcal{O}_\star}b^{(n)}_{\mathcal{O}_\star\hat\phi}$.

\item The second family of contributions, proportional to the $c^{(n)}_{q}$ coefficients, corresponds instead to the exchanges
\begin{equation} 
\mathrm{(II)}\,\, \Longrightarrow \,\,\,
    \begin{aligned}
\hat{h}_1=h_{\Phi}+\hat{h}_{\hat{\phi}}+q+l \quad &\Longrightarrow \,\,\,\, \widehat{\mathcal{O}}_1=[\hat{\phi}\Phi]_{q,l,s}=\hat{\phi}\Box_{\parallel}^{\,q}\Box_{\perp}^{\,l} \partial^{i_1}_{\perp}\dots \partial^{i_s}_{\perp} \Phi + (\dots)\,,\\
\hat{h}_2=h_\Phi+l  \quad  &\Longrightarrow \,\,\,\, \widehat{\mathcal{O}}_2=[\Phi]_{l,s}=\Box_{\perp}^{\,l}\partial^{i_1}_{\perp}\dots \partial^{i_s}_{\perp} \Phi\,,
\end{aligned}
\label{eq:exchanged_family_2}
\end{equation}
and their $1\leftrightarrow 2$ counterpart, where one of the exchanges is now a double-twist operator $[\hat{\phi}\Phi]_{q,l,s}$ made of a transverse-derivative operator $[\Phi]_{l,s}$ and the defect operator $\hat\phi$. 
The dots in~\eqref{eq:exchanged_family_2} stand for the terms in which some of the parallel derivatives act on $\hat\phi$ rather than on $\Phi$, with relative coefficients fixed by requiring $[\hat{\phi}\Phi]_{q,l,s}$ to be a primary of the parallel conformal algebra, exactly as for double-twist operators in a homogeneous CFT~\cite{Fitzpatrick:2011dm}; the transverse derivatives act on $\Phi$ only, since $\hat\phi$ has no dependence on the transverse coordinates.
These double-twists are operators also present in GFF, where they do not couple to the bulk operator $\Phi$, i.e. $b_{\Phi[\hat\phi\Phi]_{q,l,s}}^{\text{GFF}}=0$. 
The three-point coefficients $\hat{\lambda}_{[\hat\phi\Phi]_{q,l',s}[\Phi]_{l,s}\hat\phi}^{\text{GFF}}$ are however non-vanishing iff $l'=l$, which is the reason why we attribute all extra $q$ boxes in the double-twists to parallel directions: the contribution we are analyzing has to be a first correction to GFF where the bulk-to-defect coefficients $b_{\Phi[\hat\phi\Phi]_{q,l,s}}$ are now turned on.
Proceeding as we did for the first family---expanding the binomial and hypergeometric functions with~\eqref{eq:hypergeometric_and_binomial_expansion}---we extract the coefficients
\begin{equation}
    C_{[\hat{\phi}\Phi]_{q,l,s},\,[\Phi]_{l,s}}=\sum_{n}\lambda_{\Phi\Phi\mathcal{O}_\star}\, b^{(n)}_{\mathcal{O}_\star\hat\phi}\,\frac{\Gamma(2\bar{h}_\star)\Gamma(-\hat{h}_{\hat{\phi}}-q)}{\Gamma(\bar{h}_\star)\Gamma(\bar{h}_\star-\hat{h}_{\hat{\phi}}-q)}\,c^{(n)}_{q}\,P_{l}(\hat{h}_{\hat{\phi}}+q)\,.
    \label{eq:C-family-II}
\end{equation} 
This expression can now be combined with the GFF OPE coefficients of two unequal scalars and their scalar double-twist~\cite[Eq.42]{Fitzpatrick:2011dm}
\begin{equation}
    \hat{\lambda}_{[\hat\phi\Phi]_{q,l,s}[\Phi]_{l,s}\hat\phi}^{\text{GFF}} \stackrel{s\to\infty}{\simeq} \left(\frac{(2\hat{h}_{\hat{\phi}})_q (2\hat{h}_{\hat{\phi}}-\frac{p}{2}+1)_q}{q! \left(\frac{p}{2}\right)_q}\right)^{\frac12}
\end{equation}
as well as the bulk-to-defect couplings $b_{\Phi[\Phi]_{r,s}}^{\text{GFF}}$ determined in~\eqref{eq:opecoeffbulkbulk}, to get, from the definition~\eqref{eq:zbar-part-solved} of the $C_{\widehat{\mathcal{O}}_1\widehat{\mathcal{O}}_2}$ coefficients,
\begin{equation}
    b_{\Phi[\hat\phi\Phi]_{q,l,s}}\stackrel{s\to\infty}{\simeq}\frac{C_{[\hat{\phi}\Phi]_{q,l,s},\,[\Phi]_{l,s}}}{b_{\Phi[\Phi]_{l,s}}^{\text{GFF}}\,\hat{\lambda}_{[\hat\phi\Phi]_{q,l,s}[\Phi]_{l,s}\hat\phi}^{\text{GFF}}}\,\frac{s^{2h_\Phi-h_\star-1}}{\Gamma(2h_\Phi-h_\star)}\,,
    \label{eq:b-family-II-general}
\end{equation}
that is, explicitly,
\begin{equation}
    \begin{aligned}
    b_{\Phi[\hat\phi\Phi]_{q,l,s}}\stackrel{s\to\infty}{\simeq}\;&\frac{\Gamma(2h_\Phi)^{\frac12}}{\Gamma(2h_\Phi-h_\star)}\,s^{h_\Phi-h_\star-\frac12}\left[\frac{q!\left(\frac{p}{2}\right)_q}{\binom{l-\frac{d}{2}+2h_\Phi}{l}(2\hat{h}_{\hat{\phi}})_q(2\hat{h}_{\hat{\phi}}-\frac{p}{2}+1)_q}\right]^{\frac12}\\
    &\hspace{30pt}\times\sum_{n}\lambda_{\Phi\Phi\mathcal{O}_\star}\, b^{(n)}_{\mathcal{O}_\star\hat\phi}\,\frac{\Gamma(2\bar{h}_\star)\Gamma(-\hat{h}_{\hat{\phi}}-q)}{\Gamma(\bar{h}_\star)\Gamma(\bar{h}_\star-\hat{h}_{\hat{\phi}}-q)}\,c^{(n)}_{q}\,P_{l}(\hat{h}_{\hat{\phi}}+q)\,.
    \end{aligned}
    \label{eq:b-family-II}
\end{equation}
As for the three-point couplings of the first family, the bulk-to-defect couplings of the double-twist operators are suppressed by a factor $s^{-h_\star}$ with respect to the GFF couplings $b_{\Phi[\Phi]_{l,s}}^{\text{GFF}}\sim s^{h_\Phi-\frac12}$ of the transverse derivative operators, and are proportional to the bulk CFT data $\lambda_{\Phi\Phi\mathcal{O}_\star}b^{(n)}_{\mathcal{O}_\star\hat\phi}$.
\end{itemize}

This completes our discussion for generic CFTs. 
In the next section we apply the machinery developed here to a specific correlator in $\mathcal{N}=4$ SYM in the presence of a line defect.

\section{Application: \texorpdfstring{line defect in $\mathcal{N}=4$}{N=4} SYM}
\label{section:Applications}
We now present a concrete application of the bootstrap technology we developed above. Let us consider four-dimensional $\mathcal{N}=4$ SYM in the presence of a superconformal line defect, e.g. a $1/2$-BPS Maldacena-Wilson line, defined as \cite{Maldacena:1998im}
\begin{equation}
    \mathcal{W}_{\ell}:=\frac{1}{N}\mathrm{Tr}\,\mathcal{P}\,\exp{\int_{-\infty}^{+\infty}\,\mathrm{d}\tau \,(i\dot{x}_{\mu}A^{\mu}+|\dot{x}|\theta_{M}\phi^{M})}\,,
\end{equation}
where $A_{\mu}=T_{a}A^{a}_{\mu}$ and $\phi^{M}=T_{a}\phi_{a}^{M}$ are respectively the gauge and scalar fields ($T_{a}$ are the generators of the gauge group $SU(N)$ of SYM) and $M=1,\ldots,6$ is an $\mathfrak{so}(6)_{R}$ $R$-symmetry index.
Here, the line $\ell$ is parametrized by $x^{\mu}(\tau)$, with orientation chosen to be $\dot{x}^{\mu}(\tau)=(1,0,0,0)$, while the unit vector $\theta^{M}=(0,0,0,0,0,1)$ selects the scalar that couples to the line.

We consider then the $1/2$-BPS chiral primary operator
\begin{equation}
    \mathcal{O}_{20'}(x,Y):=\mathrm{Tr}\left[\phi^{M}(x)\phi^{N}(x) Y_M Y_N \right]\,,
\end{equation}
where $\phi^M$ are again the six scalar fields in $\mathcal{N}=4$ SYM and $Y_M$ are complex null polarization vectors ($Y\cdot Y=0$) that project on the symmetric traceless representation ($\mathbf{20'}$ or $[0,2,0]$) of the $\mathfrak{so}(6)$ $R$-symmetry algebra.
This operator has protected scaling dimension $\Delta=2$ and is the superprimary of the stress tensor supermultiplet. Its two-point function in the presence of a line defect has been extensively studied in the literature \cite{Barrat:2020vch,Barrat:2021yvp,Gimenez-Grau:2023fcy}. Here we would like to access a new class of CFT data by adding an operator on the defect. We choose the simplest possible defect scalar: the tilt operator.

The tilt operator is present in the spectrum of any conformal defect which breaks a global symmetry of the bulk CFT. In particular, if the bulk theory has a global symmetry with a current $J^{\mu}(x)$, the breaking of such symmetry is encoded in the equation
\begin{equation}
   \partial_{\mu}J^{\mu}(x)=\delta^q(x_{\perp})  \hat{\phi}(x_{\parallel})
\end{equation}
where the operator $\hat{\phi}$ has protected scaling dimension $\hat{\Delta}=p$ (in particular $\hat{\Delta}=1$ for line defects). A line defect in $\mathcal{N}=4$ SYM breaks the $\mathfrak{psu}(2,2|4)$ superconformal algebra down to $\mathfrak{osp}(4^*|4)\supset\mathfrak{sl}(2,\mathbb{R})\oplus\mathfrak{so}(3)\oplus\mathfrak{so}(5)_{R}$. The preserved R-symmetry algebra is $\mathfrak{so}(5)\subset\mathfrak{so}(6)$ and we have five tilt operators, associated to the five broken generators, transforming in the vector representation of $\mathfrak{so}(5)$
\begin{equation}
    \hat{\phi}(x_{\parallel},y)=\phi^m y_m
\end{equation}
where $m=1,\dots,5$ and $y_m$ in this case is a $\mathfrak{so}(5)$ null polarization vector. This operator is also the superprimary of the displacement supermultiplet.

We are interested in the three-point function
\begin{equation}
    \expval{\mathcal{O}_{20'}(x_1,Y_1)\mathcal{O}_{20'}(x_2,Y_2)\hat{\phi}(x_{\parallel},y)}
    \label{eq:N=4_correlator}
\end{equation}
To carry out a complete analysis of this correlator one would need to study the constraints of supersymmetry finding superconformal Ward identities and superconformal blocks, a task which we leave for future work. Here we only exploit the knowledge of the bulk OPE selection rules to extract information about the large-spin defect spectrum.

The superselection rules for the fusion of two chiral primary operators are well known and, in the notation of \cite{Dolan:2002zh}, they read
\begin{equation}
\begin{split}
    \mathcal{B}_{[0,2,0](0,0)}^{\frac{1}{2},\frac{1}{2}}\times\mathcal{B}_{[0,2,0](0,0)}^{\frac{1}{2},\frac{1}{2}}\to &\,\mathbf{1}+\mathcal{B}_{[0,2,0](0,0)}^{\frac{1}{2},\frac{1}{2}}+\mathcal{B}_{[0,4,0](0,0)}^{\frac{1}{2},\frac{1}{2}}+\mathcal{B}_{[2,0,2](0,0)}^{\frac{1}{4},\frac{1}{4}}\\
    & +\sum_{l=0}^{\infty}\mathcal{C}_{[0,2,0](j,j)}^{\frac{1}{2},\frac{1}{2}}+\sum_{l=0}^{\infty}\mathcal{C}_{[1,0,1](j,j)}^{\frac{1}{4},\frac{1}{4}}+\sum_{\Delta,l}\mathcal{A}^{\Delta}_{[0,0,0](j,j)}\,,
    \end{split}
\end{equation}
with $J=2j$. $\mathcal{B}^{\frac{1}{2},\frac{1}{2}}$ and $\mathcal{B}^{\frac{1}{4},\frac{1}{4}}$ represent $1/2$-BPS and $1/4$-BPS short supermultiplets respectively. The $\mathcal{C}$ supermultiplets are semishort, while the $\mathcal{A}$ labels long multiplets.
For the case of the bulk two-point function analyzed in \cite{Barrat:2020vch,Barrat:2021yvp,Gimenez-Grau:2023fcy}, one needs to select only the supermultiplets on the r.h.s. that have a non-vanishing one-point function in the presence of the conformal defect, i.e. the identity operator $\mathds{1}$, the $1/2$-BPS short multiplet $\mathcal{B}_{[0,2k,0]}$ with $k=1,2$, the semishort multiplets $\mathcal{C}_{[0,2,0],(j,j)}$ and long multiplets $\mathcal{A}_{[0,0,0],(j,j)}^{\Delta}$ with $\Delta\ge J+2$. In the lightcone bootstrap analysis, as we discussed earlier in this work, the leading contribution in the bulk channel comes from the lowest-twist exchanged operator, which in that case is just the identity operator.

For the BBD correlator we need to select all the supermultiplets that have a non-vanishing bulk-defect two-point function with the tilt operator inserted on the defect. A simple representation theory analysis shows that the allowed supermultiplets are
\begin{equation}
\begin{split}
    \mathcal{B}_{[0,2,0](0,0)}^{\frac{1}{2},\frac{1}{2}}\times\mathcal{B}_{[0,2,0](0,0)}^{\frac{1}{2},\frac{1}{2}}\to &\,\mathcal{B}_{[0,2,0](0,0)}^{\frac{1}{2},\frac{1}{2}}+\mathcal{B}_{[0,4,0](0,0)}^{\frac{1}{2},\frac{1}{2}} +\sum_{l=0}^{\infty}\mathcal{C}_{[0,2,0](j,j)}^{\frac{1}{2},\frac{1}{2}}\\
    &+\sum_{l=0}^{\infty}\mathcal{C}_{[1,0,1](j,j)}^{\frac{1}{4},\frac{1}{4}}+\sum_{\Delta,l}\mathcal{A}^{\Delta}_{[0,0,0](j,j)}\,.
    \end{split}
\end{equation}
We are interested in isolating the lowest-twist contribution in this OPE. Since the action of the supercharges cannot lower the twist of an operator, we only need to analyze the twist of the superprimary in each multiplet. The superprimary of the $1/2$-BPS multiplet $\mathcal{B}_{[0,2,0],(0,0)}^{\frac{1}{2},\frac{1}{2}}$, i.e. the stress-tensor multiplet, is the chiral primary $\mathcal{O}_{20'}$ itself and has twist $2h=2$. The next short multiplet $\mathcal{B}_{[0,4,0],(0,0)}^{\frac{1}{2},\frac{1}{2}}$ contains a scalar superconformal primary whose conformal dimension is $\Delta=4$, therefore it is subleading in twist. Both semishort multiplets $\mathcal{C}_{[0,2,0](j,j)}^{\frac{1}{2},\frac{1}{2}}$ and $\mathcal{C}_{[1,0,1](j,j)}^{\frac{1}{4},\frac{1}{4}}$ contain a superconformal primary with conformal dimension $\Delta=4+2j$, i.e. subleading twist $2h=4$. Finally, for the long multiplet $\mathcal{A}^{\Delta}_{[0,0,0](j,j)}$ the unitarity bound imposes that $\Delta\ge2+J$. Therefore, the only case we need to worry about is the threshold when the bound is saturated. In that case the long multiplet recombines in terms of shorter multiplets \cite{Dolan:2002zh}
\begin{equation}
\begin{split}
    & \mathcal{A}^{2}_{[0,0,0](0,0)}\simeq\mathcal{C}_{[0,0,0](0,0)}^{1,1}+\mathcal{B}_{[2,0,2](0,0)}^{\frac{1}{4},\frac{1}{4}}+\mathcal{D}_{[2,0,0](0,0)}^{\frac{1}{4},\frac{3}{4}}+\mathcal{D}_{[0,0,2](0,0)}^{\frac{3}{4},\frac{1}{4}}\,,\\
    & \mathcal{A}^{2+2j}_{[0,0,0](j,j)}\simeq\mathcal{C}_{[0,0,0](j,j)}^{1,1}+\mathcal{C}_{[1,0,1](j-\frac{1}{2},j-\frac{1}{2})}^{\frac{1}{4},\frac{1}{4}}+\mathcal{C}_{[1,0,0](j-\frac{1}{2},j)}^{\frac{1}{4},\frac{3}{4}}+\mathcal{C}_{[0,0,1](j,j-\frac{1}{2})}^{\frac{3}{4},\frac{1}{4}}\,.
    \end{split}
\end{equation}
The superprimary is then contained in the $\mathcal{C}_{[0,0,0](j,j)}^{1,1}$ multiplet. These multiplets however are absent in an interacting theory because they contain higher-spin conserved currents~\cite{Maldacena:2011jn}. In other words, we do not expect long multiplets to hit the unitarity bound in an interacting theory. We can then conclude that the leading-twist contribution is given by the stress tensor supermultiplet.

The twist-two operators in the stress tensor multiplet are: the scalar $\mathcal{O}_{20'}$ with $\Delta=2$, the R-symmetry current $\mathcal{J}_{\mu}$ with $\Delta=3$ and $J=1$ and the stress-tensor $T_{\mu\nu}$ with $\Delta=4$ and $J=2$. 
The stress tensor exchange can actually be ruled out in the correlator~\eqref{eq:N=4_correlator}, as it is an R-symmetry singlet which does not couple to the tilt operator.
In the following, we show explicitly how lightcone bootstrap techniques apply to the case of the dimension-two scalar, but the R-symmetry current case is completely analogous, see footnote~\ref{footnote_antisymmetric}. 
Furthermore, we expect supersymmetric Ward identities to provide an explicit relation between $b_{\mathcal{O}\hat{\phi}}$ and $b_{\mathcal{J}_{\mu} \hat{\phi}}$, but we leave this analysis for the future.

We start from the simplified crossing equation in the lightcone limit~\eqref{eq:crossing_simplified} and we fix conformal dimensions and spins for all the external operators and the lowest-twist bulk exchanged operator. 
In particular, we set $n=0$ (the only allowed tensor structure) and fix $\bar{h}_{\star}=h_{\Phi}=h_{\mathcal{O}_{20'}}=1$, $\hat{h}_{\hat{\phi}}=\frac{1}{2}$, and $d=4$. The result admits a simple closed form
\begin{equation}
    2\lambda_{\mathcal{O}_{20'}}b_{\mathcal{O}_{20'}\hat{\phi}}\,\frac{z}{1-z}\left(\arccot{(x)}-\arccot{\left(\frac{x}{\sqrt{z}}\right)}\right)
    =\sum_{\hat{h}_1,\hat{h}_2} C_{\widehat{\mathcal{O}}_1\widehat{\mathcal{O}}_2}\,\left(x^2\right)^{\hat{h}_{1}}\left(\frac{z}{x^2}\right)^{\hat{h}_{2}}\,,
    \label{eq:LHS_N=4}
\end{equation}
where we defined $\lambda_{\mathcal{O}_{20'}}\equiv\lambda_{\mathcal{O}_{20'}\mathcal{O}_{20'}\mathcal{O}_{20'}}$.
Expanding the LHS in the regime $z\ll x^2 \ll 1$, we get
\begin{equation}
        2\lambda_{\mathcal{O}_{20'}}b_{\mathcal{O}_{20'}\hat{\phi}}\sum_{l=0}^{\infty}z^{l+1}\left[\frac{\pi}{2}- \left(\sum_{k=0}^{\infty}(-1)^{k}\frac{x^{2k+1}}{2k+1}+\sum_{j=0}^{\infty}(-1)^{j}\frac{\left(\frac{z}{x^2}\right)^{j+\frac{1}{2}}}{2j+1}\right)\right].
        \label{eq:N=4_LHS_expanded}
\end{equation}
Here we can recognize the two families discussed in Section~\ref{sec:general-results}:
\begin{itemize}
    \item Family (I) defined in~\eqref{eq:exchanged_family_1} corresponds to the term proportional to $\pi/2$. In this case, the power matching enforces $\hat{h}_{1}=\hat{h}_{2}=l+1$, i.e. the transverse derivative operators exchanged in the defect channel have to be identical. The explicit matching of the two sides leads to the coefficients
    \begin{equation}
    C_{[\mathcal{O}]_{l+m,s}[\mathcal{O}]_{l,s}}=\pi\lambda_{\mathcal{O}_{20'}}b_{\mathcal{O}_{20'}\hat{\phi}}\delta_{m,0}\,.
    \end{equation}
    This is precisely what one obtains from the general expression~\eqref{eq:C-family-I} once specialized to the quantum numbers of the correlator we are considering here.
    Similarly, specializing~\eqref{eq:lambda-family-I} to the same values, the large-spin three-point couplings of the tilt with two transverse-derivative operators are
    \begin{equation}
    \hat{\lambda}_{[\mathcal{O}]_{l+m,s}[\mathcal{O}]_{l,s}\hat\phi}\stackrel{s\to\infty}{\simeq}\frac{\pi\lambda_{\mathcal{O}_{20'}}b_{\mathcal{O}_{20'}\hat{\phi}}}{s}\delta_{m,0}\,.
    \label{eq:lambda-family-I-N=4}
    \end{equation}
    The diagonal couplings decay as $s^{-h_\star}=s^{-1}$, as expected from the general analysis, while the off-diagonal ones are not generated by the twist-two exchange and are therefore suppressed by additional powers of $1/s$, controlled by higher-twist bulk operators.
    \item Family (II) defined in~\eqref{eq:exchanged_family_2} corresponds to the object in round brackets in~\eqref{eq:N=4_LHS_expanded}, where the two different terms are related by $x^2\to z/x^2$ at fixed $z$, i.e. the exchange of the two bulk operators. These terms imply $\hat{h}_{2}=1+l$, and $\hat{h}_{1}=\hat{h}_{2}+q+\frac{1}{2}=\frac{3}{2}+q+l$ and its $1\leftrightarrow 2$ counterpart, i.e. the exchange of a transverse derivative operator $\left[\mathcal{O}\right]_{l,s}$ and its double-twist with the tilt operator $\bigl[\hat\phi\mathcal{O}\bigr]_{q,l,s}$. The explicit matching leads to the coefficients
    \begin{equation}
    C_{[\hat{\phi}\mathcal{O}]_{q,l,s}[\mathcal{O}]_{l,s}}=\frac{2\lambda_{\mathcal{O}_{20'}}b_{\mathcal{O}_{20'}\hat{\phi}}(-1)^{q+1}}{2q+1}\,,
    \end{equation}
    which again follows from the general expression~\eqref{eq:C-family-II}.
    The bulk-to-defect couplings of the double-twist operators then follow from~\eqref{eq:b-family-II}. For a line defect, $p=1$, with $\hat h_{\hat\phi}=\tfrac12$ the GFF three-point coefficients reduce to $\hat{\lambda}^{\text{GFF}}_{[\hat\phi\mathcal{O}]_{q,l,s}[\mathcal{O}]_{l,s}\hat\phi}\simeq(2q+1)^{\frac12}$, while $b^{\text{GFF}}_{\mathcal{O}[\mathcal{O}]_{l,s}}\simeq s^{\frac12}$, so that
    \begin{equation}
    b_{\mathcal{O}_{20'}[\hat{\phi}\mathcal{O}]_{q,l,s}}\stackrel{s\to\infty}{\simeq}\frac{2\lambda_{\mathcal{O}_{20'}}b_{\mathcal{O}_{20'}\hat{\phi}}\,(-1)^{q+1}}{(2q+1)^{\frac32}}\,s^{-\frac12}\,,
    \label{eq:b-family-II-N=4}
    \end{equation}
    for any $l$. These couplings are suppressed by $s^{-h_\star}=s^{-1}$ with respect to the GFF couplings of the transverse-derivative operators, and alternate in sign with the number $q$ of parallel derivatives.
\end{itemize}
We have therefore seen how the leading-twist exchange of $\mathcal{O}_{20'}$ fixes two new sets of defect CFT data: the OPE coefficients $\hat{\lambda}_{[\mathcal{O}]_{l+m,s}[\mathcal{O}]_{l,s}\hat\phi}$ and the bulk-to-defect coefficients $b_{\mathcal{O}_{20'}[\hat{\phi}\mathcal{O}]_{q,l,s}}$ at large transverse spin. 
Note that the answer presented here focused only on the $\mathcal{O}_{20'}$ exchange, and thus contributes only to the R-symmetry structure associated with this operator exchange; the $\mathcal{J}_\mu$ R-symmetry current contribution is expected to be related to the result presented here via superconformal Ward identities, and must contribute to a different R-symmetry channel, one that is antisymmetric under the exchange of the two external bulk operators.
We finally stress that, while the results of this section are framed in the context of the $1/2$-BPS Maldacena-Wilson line in $\mathcal{N}=4$ SYM, their derivation only relied on the dimensions of the operators involved.
The contribution computed here is therefore universal: in any four-dimensional theory with a line defect, a bulk scalar of dimension two and a defect scalar of dimension one, the exchange of a twist-two scalar in the bulk channel produces exactly the defect data~\eqref{eq:lambda-family-I-N=4} and~\eqref{eq:b-family-II-N=4}. In theories where other operators share the leading twist, for instance the stress tensor whenever it couples to the external defect scalar, their contributions simply add to the ones obtained here, as discussed in Section~\ref{section:Multipoint_dCFT}.

\section{Outlook}
\label{section:Outlook}
In this paper we focused on correlators involving two bulk operators and one defect operator, constraining the large-transverse-spin OPE coefficients $\hat{\lambda}_{[\Phi]_{l+m,s}[\Phi]_{l,s}\hat\phi}$ of two transverse derivative operators and a generic defect operator, as well as the bulk-to-defect couplings $b_{\Phi[\hat\phi\Phi]_{q,l,s}}$ involving defect double-twist operators.  
The main results for general CFTs are the closed-form expressions~\eqref{eq:lambda-family-I} and~\eqref{eq:b-family-II}, while those specific to some correlators in $d=4$ theories such as $\mathcal{N}=4$ SYM are~\eqref{eq:lambda-family-I-N=4} and~\eqref{eq:b-family-II-N=4}.
Both sets of data correspond to first corrections to GFF, suppressed as $s^{-h_\star}$ with power dictated by the leading twist of the theory.
The straightforward success of this approach leads to some natural directions for future exploration. 

The appealing feature of the setup we considered here is the simplicity compared to higher-point correlators without defects. The insertion of operators on the defect affects the kinematics in a controlled way. For instance, the defect-channel conformal blocks preserve the factorized structure that was crucial in the derivation of the Lorentzian inversion formula \cite{Lemos:2017vnx} and the dispersion relation \cite{Bianchi:2022ppi,Barrat:2022psm}. It is therefore natural to ask whether these techniques can be extended to our setup. This would give a rigorous and systematic way to understand the large-spin regime of the theory and it would provide a powerful computational technique for those theories that admit a perturbative expansion with some control over the spectrum of exchanged operators.

Throughout the paper we considered only defects with codimension greater than one. In some respect, the case of codimension-one defects (boundaries or interfaces) should be even simpler because the number of conformal cross ratios is further reduced, see for instance~\cite{Chen:2023yvw}. However, in this case there is no notion of transverse spin and the lightcone limit of the crossing equation should provide information on the defect spectrum at large conformal dimension. We leave a detailed analysis of this case for future work.

In the context of $\mathcal{N}=4$ SYM, as we hinted upon in Section~\ref{section:Applications}, an immediate future step to pursue is the determination of superconformal Ward identities for the $\langle\mathcal{O}_{20'}\mathcal{O}_{20'}\hat\phi\rangle$ correlator. This would allow for a supersymmetrization of our results, relating the $\mathcal O_{20'}$ and $\mathcal J_\mu$ contributions and thus determining the complete leading-twist answer for this correlator.  

Finally, another natural direction is the extension of lightcone bootstrap techniques to other higher-point correlators. We sketched some possibilities in Figure \ref{fig:higher-point-cases}. In particular, it would be very interesting to analyze the BBDD and the BBB case, which would give access to a whole new class of operators. 
Moreover, one may wonder whether the simplicity we observe in the structure of the defect conformal blocks survives after additional insertions, and whether some extrapolations to a large number of insertions can be made. We plan to address this in the future.

\newpage
\acknowledgments
The authors would like to thank M. Lemos, M. Meineri and V. Schomerus for useful discussions. LB's and LQ's research is partially supported
by the MUR PRIN contract 2022N9CTAE "Constraining strongly coupled quantum field theories using symmetry". LB and AM would like to thank the Isaac Newton Institute for Mathematical Sciences, Cambridge, for support and hospitality during the programme "Quantum Field Theory with Boundaries, Impurities, and Defects" where work on this paper was undertaken. This work was supported by EPSRC grant no EP/R014604/1. The authors participate in the project HORIZON-MSCA-2023-SE-01-101182937HeI.

The inconsistency in the published version of this paper was detected during a consistency check of the defect-channel conformal blocks performed with the AI assistant Claude (Anthropic; model Claude Fable 5.1), and was then traced by the authors to the parameter interchange described in the Note added. In preparing the revised version we further used the same assistant, accessed through Claude Code, to cross-check derivations, to perform numerical checks of the corrected formulas and to help draft and edit the text. All results were verified by the authors, who take full responsibility for the content.

\appendix
\section{Embedding space formalism}
\label{section:embedding}
Let us consider a $d$-dimensional Euclidean CFT living in $\mathbb{R}^{d}$, with a $p$-dimensional flat conformal defect. We work in the embedding space formalism, which has the advantage that conformal transformations act linearly on the coordinates. In particular, the embedding space realizes the $d$-dimensional flat space as a subset of the projective null cone in $\mathbb{R}^{d+1,1}$. Indices in embedding space are $A,B,\ldots=(+,-,1,\ldots,p)$ for the $SO(p+1,1)$ defect conformal group and $I,J,\ldots=(1,\ldots,d-p)$ for the transverse rotation part $SO(d-p)$. We will also use the definition $q=d-p$ to indicate the codimension.  The defect location is such that all components of the transverse indices $I,J,\ldots$ are zero.
The two indices are associated with two scalar products
\begin{equation}
    P \parsc Q = P^A \eta_{AB}Q^B\,, \qquad P\trasc Q= P^I \delta_{IJ} Q^J\,.
\end{equation}
For spinning operators, we will use the index-free notation. In particular, a symmetric traceless bulk operator $\mathcal{O}_{\Delta,J}(P,Z)$ is obtained by contracting the indices with a polarization vector $Z$ such that $P^2=0$, $Z^2=0$, $Z\cdot P=0$. Splitting in parallel and orthogonal directions we have
\begin{equation}
    P\parsc P=-P \trasc P\,, \qquad Z\parsc Z=-Z\trasc Z\,, \qquad Z\parsc P = -Z\trasc P\,.
\end{equation}
As usual, due to the transversality condition (gauge invariance), $Z$ can only appear in correlation functions through a wedge product. In this case, though, only some mixed components are independent 
\begin{equation}
    C^{AI}=(P\wedge Z)^{AI} = P^A Z^I - P^I Z^A\,.
    \label{eq:def CAI}
\end{equation}
For example, the structure of a one-point function of a primary bulk operator $\Phi_{\Delta,J}(P,Z)$ can be constructed with this tensor structure $C^{AI}$. In particular, the $Z$ polarization can only appear from $C_{AI}C^{AI}$, which implies that only even spin operators acquire a non-vanishing one-point function (at least for parity-preserving theories). Fixing homogeneities appropriately, we have
\begin{equation}
    \expval{O_{\Delta,J}(P,Z)} = a_\mathcal{O}\frac{\left(C_{AI}C^{AI}\right)^{\frac{J}{2}}}{\left(P\trasc P\right)^{\Delta+J}}\,,
\end{equation}
where the normalization $a_{\mathcal{O}}$ is part of the defect CFT data.

On the other hand, defect operators are represented with a hat $ \widehat{\mathcal{O}}_{\Delta,j,s}(P,Z,W)$  and characterized by three embedding space vectors $P^A$, $Z^A$, $W^I$ associated to spacetime, parallel spin and orthogonal spin respectively. They are constrained by
\begin{equation}
    P\parsc P=0\,, \qquad Z\parsc Z=0\,, \qquad Z\parsc P=0\,, \qquad W\trasc W=0\,.
\end{equation}

\section{Generalized hypergeometric functions}
\label{section:Hypergeometrics}
Throughout the paper, we encounter several special functions. Here we report some definitions for the reader's convenience, redirecting to~\cite{Matsumoto_2020} and~\cite{NIST:DLMF} for more details. 

The Lauricella $F_D$ function, a generalized multivariable hypergeometric function, can be defined as
\begin{equation}
    \begin{aligned}
    F_D\!\left(a,b_1,b_2,b_3,c;x_1,x_2,x_3\right)&=\frac{\Gamma(c)}{\Gamma(a)\Gamma(c-a)}\int_{0}^1\dd t\, t^{a-1}(1-t)^{c-a-1}\!\left(\prod_{i=1}^3 (1-x_i t)^{-b_i}\right)\\
    &=\sum_{(n_1,n_2,n_3)\in \mathbb{N}_0^3}\!\frac{(a)_{n_1+n_2+n_3} \prod_{i=1}^3(b_i)_{n_i}}{(c)_{n_1+n_2+n_3} \prod_{i=1}^{3}n_i!} x_1^{n_1}x_2^{n_2}x_3^{n_3}\,.
    \end{aligned}
    \label{eq:Lauricella_def}
\end{equation}
Note that, similarly to what happens in the case of the standard hypergeometric functions, if one of the $b_i$ is a negative integer, the sum over the associated $n_i$ truncates to a finite number of terms. This is precisely the case for the Lauricella $F_D$ that appears in our bulk-channel blocks in Section~\ref{section:Multipoint_dCFT}. In this case, the function becomes a finite sum of Appell $F_1$ functions, a class of hypergeometric functions with two variables that we define below.

The Appell function $F_{1}$ is defined in its series and integral representation as
\begin{equation}
    \begin{split}
        F_{1}(a,b_{1},b_{2},c,z_{1},z_{2})&=\sum_{m,n=0}^{\infty}\frac{(a)_{m+n}(b_{1})_{m}(b_{2})_{n}}{(c)_{m+n}\,m!\,n!}z_{1}^{m}z_{2}^{n}\\
        &=\frac{\Gamma(c)}{\Gamma(c-a)\,\Gamma(a)}\int_{0}^{1}\mathrm{d}t\,\frac{t^{a-1}(1-t)^{c-a-1}}{(1-t\,z_{1})^{b_{1}}(1-t\,z_{2})^{b_{2}}}\,.
        \label{eq:appellF1}
    \end{split}
\end{equation}
This function is relevant in the decomposition of the Lauricella function we encounter in the bulk channel; in particular, for the exchange of a scalar leading-twist operator the bulk-channel block of Section~\ref{section:Multipoint_dCFT} reduces to a single Appell $F_1$ function.

In the case of defect-channel conformal blocks, the general solution is expressed in terms of an Appell function $F_{4}$, which we can define as
\begin{equation}
    \begin{split}
        F_{4}(a,&b,c_{1},c_{2},z_{1},z_{2})=\sum_{m,n=0}^{\infty}\frac{(a)_{m+n}(b)_{m+n}}{(c_{1})_{m}(c_{2})_{n}\,m!\,n!}z_{1}^{m}z_{2}^{n}\\
        &=\frac{\Gamma(c_{1})\Gamma(c_{2})}{\Gamma(a)\Gamma(b)\Gamma(c_{1}-a)\Gamma(c_{2}-b)}\\
        &\times\int_{0}^{1}\int_{0}^{1}\mathrm{d}u\,\mathrm{d}v\,\frac{u^{a-1}v^{b-1}(1-u)^{c_{1}-a-1}(1-v)^{c_{2}-b-1}}{(1-z_{1}\,u)^{c_{1}+c_{2}-a-1}(1-z_{2}\,v)^{c_{1}+c_{2}-b-1}(1-z_{1}\,u-z_{2}\,v)^{a+b-c_{1}-c_{2}+1}}\,.
    \end{split}
\end{equation}
Note that the integral representation is now a non-factorized double integral.

It is also useful to recall the following reduction formula~\cite[16.16.7]{NIST:DLMF} for the $F_{4}$ :
\begin{equation}
\begin{split}
    F_{4}&\left(a,b,c_{1},c_{2},x_{1}(1-x_{2}),x_{2}(1-x_{1})\right)= \sum_{k=0}^{\infty}\frac{(a)_{k}(b)_{k}(a+b-c_{1}-c_{2}+1)_{k}}{(c_{1})_{k}(c_{2})_{k}k!}x_{1}^{k}x_{2}^{k}\,\\
    &\times {}_{2}F_{1}(a+k,b+k,c_{1}+k,x_{1}){}_{2}F_{1}(a+k,b+k,c_{2}+k,x_{2})\,.
\end{split}
\label{eq:2F1 expansion}
\end{equation}

\section{Bulk-channel conformal blocks}
\label{section:bulk-channel-blocks}
\subsection{Bulk-bulk two-point function}
In this appendix, we review the bulk conformal blocks for the case of the bulk two-point function following the argument in \cite{Billo:2016cpy}. In the bulk channel \eqref{eq:2bulk correlator},
\begin{equation}
    f(\xi,\cos{\varphi})=\xi^{-{\Delta}_{\Phi}}\sum_{\Delta,J} \lambda_{\Phi\Phi \mathcal{O}} \,a_{\mathcal{O}} f_{\Delta,J}(\xi,\cos{\varphi})\,,
\end{equation}
there is a sum over bulk primary operators appearing in the $\Phi\times\Phi$ OPE with a non-vanishing one-point function. As we explained in appendix \ref{section:embedding}, we can restrict the sum to even J.

One way to compute a conformal block is by solving an eigenvalue equation. Each bulk conformal block $f_{\Delta,J}(\xi,\cos{\varphi})$ is an eigenfunction of the quadratic Casimir operator of the full conformal group $SO(d+1,1)$. Let us define the conformal generators in embedding space as
\begin{equation}
    \mathcal{T}_{MN}^{(i)}=P^{(i)}_{M}\frac{\partial}{\partial P^{N}_{(i)}}-P_{N}^{(i)}\frac{\partial}{\partial P^{M}_{(i)}}\,,
\end{equation}
where $M,N$ are indices in the embedding space $M,N=(+,-,1,\ldots,d)$ for the full conformal group $SO(d+1,1)$ and with $i=1,2$ corresponding to the two bulk insertions. Then, the Casimir differential equation becomes
\begin{equation}
    \left(\mathcal{C}^{2}+C_{\Delta,J}\right)\frac{\xi^{-\Delta_{\Phi}}f_{\Delta,J}(\xi,\cos{\varphi})}{(P_{1}\circ P_{1})^{\Delta_{\Phi}}(P_{2}\circ P_{2})^{\Delta_{\Phi}}}=0\,,
\end{equation}
where $\mathcal{C}^{2}=\frac{1}{2}(\mathcal{T}_{MN}^{(1)}+\mathcal{T}_{MN}^{(2)})^{2}$ and $C_{\Delta,J}=\Delta(\Delta-d)+J(J+d-2)$ is the eigenvalue.
For general dimensions of the spacetime and the defect, a complicated but exact solution was provided in~\cite{Isachenkov:2018pef} in terms of Harish-Chandra functions. However, at leading order in the lightcone limit, corresponding to taking $\xi\to0$ and keeping $\cos{\varphi}$ fixed, the authors of \cite{Billo:2016cpy} already observed that the Casimir equation simplifies to a standard hypergeometric differential equation. The solution is a factorized function of $\xi$ and $\cos{\varphi}$ of the following form:
\begin{equation}
    f_{\Delta,J}(\xi,\varphi)=\xi^{\frac{\Delta-J}{2}}\left(\sin^{J}\varphi\,\,{}_{2}F_{1}\left( \frac{\Delta+J}{4},\frac{\Delta+J}{4},\frac{\Delta+J+1}{2},\sin^{2}\varphi\right)+O(\xi)\right)\,,
\end{equation}
where we see that the lightcone limit organizes in terms of operators with lowest twist $\tau=\Delta-J$. Let us express this in terms of the cross ratios $z,\bar{z}$ defined in \eqref{eq:zbarz_def} as done in~\cite{Lemos:2017vnx}. The lightcone limit corresponds now to taking the limit $\bar{z}\to1$ and we can express the bulk conformal block as in \eqref{eq:bulkbulklightcone} with
\begin{equation}
    \tilde{f}_{\Delta+J}(z)= 2^{-J}(1-z)^{\frac{\Delta+J}{2}}\,{}_{2}F_{1}\left( \frac{\Delta+J}{2},\frac{\Delta+J}{2},\Delta+J,1-z\right)\,.
    \label{eq:bulkbulkblocklightcone}
\end{equation}

\subsection{Bulk-bulk-defect three-point function}
Let us now consider the bulk channel for the bulk-bulk-defect setup.
In the bulk channel, we can use the $P_1\cdot P_2\to 0$ lightcone OPE~\cite{Ferrara:1971zy,Ferrara:1971tq,Ferrara:1972cq} to decompose the correlator in terms of bulk-to-defect correlators. Using the notation of~\cite[Appendix~A]{Kaviraj:2022wbw} up to normalization of embedding space vectors, the scalar-scalar lightcone OPE can be written as

\begin{equation}
    \Phi_1\!(P_1)\Phi_2(P_2) \! \stackrel{P_{12}\rightarrow 0}{\simeq} \sum_{\mathcal{O}}  \!\frac{(-2P_{12})^{h_{\mathcal{O}}-h_1-h_2}\lambda_{\Phi_1\Phi_2\mathcal{O}}}{\mathrm{B}(\bar h_{\mathcal{O}}+h_{12},\bar h_{\mathcal{O}}+h_{21})} \!\int_{0}^1\!\! \frac{\dd t}{t^{1-\bar h_{\mathcal{O}}+h_{12}}} (1-t)^{h_{12}+\bar{h}_{\mathcal{O}}-1} \mathcal{O}(\tilde{P},\tilde{Z})+\dots
    \label{eq:lightcone-OPE appendix}
\end{equation}
with $\tilde{P}=P_2-t(P_2-P_1)$, $\tilde{Z}=P_2-P_1$.
This equation is compatible with the normalization of two- and three-point functions.

Using the lightcone OPE~\eqref{eq:lightcone-OPE appendix}, we can work out an expression for the lightcone bulk conformal blocks for the Bulk-Bulk-Defect correlator we are interested in. Inserting this lightcone OPE formula in our correlation function, we obtain an integral of a bulk-to-defect two-point function
\begin{equation}
    \expval{\!\mathcal{O}_{\Delta,J}(\tilde{P},\tilde{Z}) \hat{\phi}(P_3)\!}=\frac{\left(4\tilde{C}_1^{AI}P^3_A  \tilde{P}_I\right)^{J}}{\left(-2 \tilde{P}\parsc P_3\right)^{\hat\Delta_{\hat\phi}+J}\!\left(\tilde{P}\trasc \tilde{P}\right)^{\frac{\Delta-\hat\Delta_{\hat\phi} +J}{2}}} \sum_n b_{\mathcal{O}\hat\phi}^{(n)}\!\left(\!\frac{\tilde{C}^{AI} \tilde{C}_{AI}\left(\tilde{P}\parsc P_3\right)^2}{2\left(\tilde{C}^{AI}P^3_A  \tilde{P}_I\right)^2}\!\right)^{\lfloor\frac{J}{2}\rfloor-n}\hspace*{-5pt}\,,
\end{equation}
where $n=0,1,\dots,\lfloor J/2\rfloor$. In the following we take $J$ to be even, as appropriate for the OPE of two identical scalars, so that the structure linear in the polarization appears with power $2n$; for odd $J$ this power becomes $2n+1$ and the derivation below goes through with the replacement $2n\to2n+1$, see footnote~\ref{footnote_antisymmetric}. With $\tilde P_1=P_2-t(P_2-P_1)$ and $\tilde Z_1= P_2-P_1$ this becomes
\begin{equation}
    \begin{split}
       \sum_n &b^{(n)}_{\mathcal{O}\hat\phi}\frac{\left[4\left(P_1\parsc P_3\right)\left(P_2\trasc P_2\right)-4\left(P_2\parsc P_3\right)\left(P_1\trasc P_2\right)\right]^{2n}}{\left(-2P_2\parsc P_3\right)^{\hat \Delta_{\hat{\phi}}+2n}\left(P_2\trasc P_2\right)^{\frac{\Delta-\hat\Delta_{\hat{\phi}}+J}{2}}} \left[\left(P_1\trasc P_2\right)^2-\left(P_1\trasc P_1\right)\left(P_2\trasc P_2\right)\right]^{\frac{J}{2}-n}\\
    & \times
    \left[1-t\left(1-\frac{\left(P_2\parsc P_3\right)\left(P_1\trasc P_1\right)-\left(P_1\parsc P_3\right)\left(P_1\trasc P_2\right)}{\left(P_2\parsc P_3\right)\left(P_1\trasc P_2\right)-\left(P_1\parsc P_3\right)\left(P_2\trasc P_2\right)}\right)\right]^{2n}
    \left[1-t\left(1-\frac{P_1\parsc P_3}{P_2\parsc P_3}\right)\right]^{-\hat\Delta_{\hat{\phi}}-2n}\\
    & \times\left[1-t\frac{\left(P_2\trasc P_2\right)-\left(P_1\trasc P_2\right)-\sqrt{\left(P_1\trasc P_2\right)^2-\left(P_1\trasc P_1\right)\left(P_2\trasc P_2\right)}}{P_2\trasc P_2}\right]^{\frac{\hat\Delta_{\hat{\phi}}-\Delta-J}{2}}\\
    & \times
    \left[1-t\frac{\left(P_2\trasc P_2\right)-\left(P_1\trasc P_2\right)+\sqrt{\left(P_1\trasc P_2\right)^2-\left(P_1\trasc P_1\right)\left(P_2\trasc P_2\right)}}{P_2\trasc P_2}\right]^{\frac{\hat\Delta_{\hat{\phi}}-\Delta-J}{2}}\,. 
    \end{split}
\end{equation}

The square root can be evaluated as
\begin{equation}
    \sqrt{\left(P_1\trasc P_2\right)^2-\left(P_1\trasc P_1\right)\left(P_2\trasc P_2\right)}=\sqrt{\left[\left(P_1-P_2\right)\trasc P_2\right]^2}=\left(P_1-P_2\right)\trasc P_2\,,
\end{equation}
valid as long as $(P_1-P_2)\trasc(P_1-P_2)=0$ and where the sign of the square root is chosen compatibly with the standard conformal frame we consider. Then, the bulk-defect two-point function becomes
\begin{equation}
    \begin{split}
         \sum_n b^{(n)}_{\mathcal{O}\hat\phi}&\frac{\left[4\left(P_1\parsc P_3\right)\left(P_2\trasc P_2\right)-4\left(P_2\parsc P_3\right)\left(P_1\trasc P_2\right)\right]^{2n}}{\left(-2P_2\parsc P_3\right)^{\hat \Delta_{\hat{\phi}}+J-2n}\left(P_2\trasc P_2\right)^{\frac{\Delta+J-\hat\Delta_{\hat{\phi}}}{2}}} \left[\left(P_1\trasc P_2\right)^2-\left(P_1\trasc P_1\right)\left(P_2\trasc P_2\right)\right]^{\frac{J}{2}-n}\\
    & \times
    \left[1-t\left(1-\frac{\left(P_2\parsc P_3\right)\left(P_1\trasc P_1\right)-\left(P_1\parsc P_3\right)\left(P_1\trasc P_2\right)}{\left(P_2\parsc P_3\right)\left(P_1\trasc P_2\right)-\left(P_1\parsc P_3\right)\left(P_2\trasc P_2\right)}\right)\right]^{2n}\\
    &\times
    \left[1-t\left(1-\frac{P_1\parsc P_3}{P_2\parsc P_3}\right)\right]^{-\hat\Delta_{\hat{\phi}}-2n}\left[1-t\left(2-2\frac{P_1\trasc P_2}{P_2\trasc P_2}\right)\right]^{\frac{\hat\Delta_{\hat{\phi}}-\Delta-J}{2}}.
    \end{split}
\end{equation}
If we insert this in the lightcone integral, we get the integral representation for a Lauricella $F_D$ function as defined in \eqref{eq:Lauricella_def}. This allows us to express the correlator expanded in the bulk channel as
\begin{equation}
    \begin{split}
        \langle\Phi(P_1)\Phi(P_2)&\hat\phi(P_3)\rangle=
    \!(-2P_{12})^{h-2h_{\Phi}}\lambda_{\Phi\Phi\mathcal{O}}\!\sum_n b^{(n)}_{\mathcal{O}\hat\phi}\frac{\left[4\!\left(P_1\parsc P_3\right)\left(P_2\trasc P_2\right)\!-\!4\!\left(P_2\parsc P_3\right)\left(P_1\trasc P_2\right)\right]^{2n}}{\left(-2P_2\parsc P_3\right)^{\hat \Delta_{\hat{\phi}}+J-2n}\left(P_2\trasc P_2\right)^{h+J-\frac{\hat\Delta_{\hat{\phi}}}{2}}}\\ &\times\left[\left(P_1\trasc P_2\right)^2-\left(P_1\trasc P_1\right)\left(P_2\trasc P_2\right)\right]^{\frac{J}{2}-n}
    F_D\!\left(a,b_-,b_0,b_+,c;x_-,x_0,x_+\right)\,,
    \label{eq:bulk-channel-LC-embedding}
    \end{split}
\end{equation}
where
\begin{equation}
    \begin{gathered}
    x_-=1-\frac{\left(P_2\parsc P_3\right)\left(P_1\trasc P_1\right)-\left(P_1\parsc P_3\right)\left(P_1\trasc P_2\right)}{\left(P_2\parsc P_3\right)\left(P_1\trasc P_2\right)-\left(P_1\parsc P_3\right)\left(P_2\trasc P_2\right)}\,, \quad x_0=2-2\frac{P_1\trasc P_2}{P_2\trasc P_2}\,, \quad x_+=1-\frac{P_1\parsc P_3}{P_2\parsc P_3}\,,\\
    a=h+J\,,\qquad b_-=-2n\,, \qquad b_0=h+J-\frac{\hat\Delta_{\hat{\phi}}}{2}\,,\qquad b_+=2n+\hat\Delta_{\hat{\phi}}\,,\qquad c=2(h+J)\,,
    \end{gathered}
\end{equation}
with $h=\frac{\Delta-J}{2}$ being the half twist of the bulk exchanged operators. Let us also define for simplicity $\bar{h}=h+J=\frac{\Delta+J}{2}$ and $\hat{h}_{\hat{\phi}}=\frac{\Delta_{\hat{\phi}}}{2}$.
We notice that one of the sums of the Lauricella truncates, due to $-2n$ in the second argument being a negative integer.
Fixing the positions of fields as in the conformal frame depicted in Figure~\ref{fig:BBD-frame-nolimit}, the bulk-channel expansion of the correlator~\eqref{eq:bulk-channel-LC-embedding} becomes:
\begin{multline}
    \langle\Phi(P_1)\Phi(P_2)\hat\phi(P_3)\rangle\stackrel{\text{frame}}{=}
    \sum_{\mathcal{O},n}\lambda_{\Phi\Phi\mathcal{O}}\, b^{(n)}_{\mathcal{O}\hat\phi}(1-\bar z)^{h-2h_\Phi}(1-z)^{\bar{h}-2h_\Phi}z^{-\bar{h}+\hat{h}_{\hat{\phi}}}\frac{(x^2-z)^{2n}}{(x^2+z)^{2n+2\hat{h}_{\hat{\phi}}}}\\
    \times F_D\!\left(\bar{h},-2n,\bar{h}-\hat{h}_{\hat{\phi}},2(\hat{h}_{\hat{\phi}}+n),2\bar{h};\frac{z-1}{z-x^2},\frac{z-1}{z},\frac{z-1}{z+x^2}\right)\,.
    \label{eq:bulk channel bbd appendix}
\end{multline}
Note that this expression, which we derived from~\eqref{eq:bulk-channel-LC-embedding}, is not manifestly symmetric under $1\leftrightarrow 2$. The lightcone OPE is, however, a symmetric operation, so there must exist Lauricella identities that allow us to express the conformal blocks with equivalent representations. One such representation is obtained by exchanging $P_1\leftrightarrow P_2$ in~\eqref{eq:bulk-channel-LC-embedding}, obtaining
\begin{multline}
    \langle\Phi(P_1)\Phi(P_2)\hat\phi(P_3)\rangle\stackrel{\text{frame}}{=}\sum_{\mathcal{O},n}\lambda_{\Phi\Phi\mathcal{O}}\, b^{(n)}_{\mathcal{O}\hat\phi}\left(1-\bar{z}\right)^{h-2 h_{\Phi }}\frac{(1-z)^{\bar{h}-2 h_{\Phi }}\left(1-x^2\right)^{2n}}{\left(x^2+1\right)^{2 \hat{h}_{\hat{\phi}}+2n}}  \\
    \times F_D\!\left(\bar{h},-2n,\bar{h}-\hat{h}_{\hat{\phi}},2( \hat{h}_{\hat{\phi}}+n),2 \bar{h},\frac{z-1}{x^2-1},1-z,\frac{1-z}{x^2+1}\right)\,.
    \label{eq:bulk channel zx}
\end{multline}

These entail non-trivial identities between Lauricella functions, which allow for analytic continuations in regimes where one of the representations is problematic.

As a further consistency check for our bulk conformal block, let us consider the case in which the external scalar defect operator $\hat{\phi}(P_{3})$ is just the identity. The expectation is to reproduce the bulk conformal block for the bulk two-point function in the lightcone limit~\eqref{eq:bulkbulkblocklightcone}. To impose the external defect operator to be the identity we need to fix $\hat{h}_{\hat{\phi}}=0$ and trivialize the tensor structure label to $n=0$.  The bulk channel in~\eqref{eq:bulk channel bbd appendix} then loses all dependence on the $x$ cross-ratio and reduces to 
\begin{equation}
    \begin{split}
       \langle\Phi(P_1)\Phi(P_2)\rangle\stackrel{\text{frame}}{=}
    \sum_{\mathcal{O}}\lambda_{\Phi\Phi\mathcal{O}}\, b_{\mathcal{O}\hat{\mathds{1}}}(1-\bar z)^{h-2h_\Phi}(1-z)^{\bar{h}-2h_\Phi}z^{-\bar{h}}\,{}_{2}F_{1}\!\left(\bar{h},\bar{h},2\bar{h};\frac{z-1}{z}\right)\,, 
    \end{split}
\end{equation}
where $b_{\mathcal{O}\hat{\mathds{1}}}=a_{\mathcal{O}}$ is just the one-point function coefficient. Using the following identity for the hypergeometric function ${}_{2}F_{1}$

\begin{equation}
    {}_{2}F_{1}\left(a,c-b,c,\frac{z}{z-1}\right)=(1-z)^{a}\,{}_{2}F_{1}(a,b,c,z),\quad z\notin(1,\infty),\,
    \label{eq:trasf2F1a}
\end{equation}
then, we can reduce the bulk channel as

\begin{equation}
    \begin{split}
       \langle\Phi(P_1)\Phi(P_2)\rangle\!&\stackrel{\text{frame}}{=}
    \!\sum_{\mathcal{O}}\!\lambda_{\Phi\Phi\mathcal{O}}\, a_{\mathcal{O}}(1-\bar z)^{\frac{\Delta-J}{2}-\Delta_{\Phi}}(1-z)^{\frac{\Delta+J}{2}-\Delta_{\Phi}}\,{}_{2}F_{1}\!\left(\tfrac{\Delta+J}{2},\tfrac{\Delta+J}{2},\Delta+J;1-z\right)\\
    &=(z\bar{z})^{-\frac{\Delta_{\Phi}}{2}}\left(\frac{(1-\bar{z})(1-z)}{(z\bar{z})^{1/2}}\right)^{-\Delta_{\Phi}}\sum_{\Delta,J}\lambda_{\Phi\Phi\mathcal{O}}\, a_{\mathcal{O}}(1-\bar{z})^{\frac{\Delta-J}{2}}2^{J}\tilde{f}_{\Delta+J}(z)\,, 
    \end{split}
\end{equation}
where on the second line we multiplied and divided for $(z\bar{z})^{-\Delta_{\Phi}/2}$. It is straightforward to notice that this expression, with $\tilde{f}_{\Delta+J}(z)$ defined in \eqref{eq:bulkbulkblocklightcone}, is equivalent\footnote{Actually, our result differs with the one in \cite{Lemos:2017vnx} for a factor $2^{-J}$, which we absorb it in the definition of the bulk-to-defect OPE coefficient $b_{\mathcal{O}\hat{\phi}}^{(n)}$.} to the bulk channel for the bulk two-point function case \eqref{eq:2bulk correlator}, \eqref{eq:bulkexp}, \eqref{eq:bulkbulklightcone} in the lightcone limit.

\section{Large spin expansion of defect-channel blocks}
\label{section:largespinexpansion}
In this appendix, we will show how to perform the large transverse spin expansion of the defect conformal blocks for the cases of the bulk-bulk two-point function and the bulk-bulk-defect three-point function.

\paragraph{Bulk-bulk blocks in the defect channel}
Let us first consider the bulk-bulk conformal block in the defect channel obtained analytically in \cite{Billo:2016cpy}, and let us write it in the convention of \cite{Lemos:2017vnx}:
\begin{equation}
\hat{f}_{\hat{h},s}(z,\bar{z})=z^{\hat{h}}\bar z^{\hat{h}+s} {}_{2}F_{1}\!\left(-s, \frac{q}{2}-1, 2-\frac{q}{2}-s; \frac{z}{\bar z}\right) {}_2F_1\!\left(2\widehat{h}+s,\frac{p}{2},2\widehat{h}+s+1-\frac{p}{2}; z \bar z\right)\,.  
\label{eq:defect block bb}
\end{equation}

In the lightcone bootstrap, one is interested in reproducing the low-twist bulk exchange in the lightcone limit $\bar{z}\rightarrow1$ in the defect channel. For the simpler case of a bulk two-point function, the lowest-twist contribution in the bulk OPE is the identity. Here, the crossing equation takes the form
\begin{equation}
    1=\lim_{\bar{z}\rightarrow1}\left(\frac{(1-z)(1-\bar{z})}{\sqrt{z\bar{z}}}\right)^{\Delta_{\Phi}}\sum_{\hat{h},s}b_{\hat{h},s}^{2}\hat{f}_{\hat{h},s}(z,\bar{z})\,.
    \label{eq:crossing bb identity}
\end{equation}
Since each conformal block is analytic in the region of interest $1-\bar{z}\ll z<1$, then in the lightcone limit $(1-\bar{z})\rightarrow0$ one finds that $(1-\bar{z})^{\Delta_{\Phi}}\hat{f}_{\hat{h},s}=0$ for fixed $\hat{h}$ and $s$, so the sum on the RHS does not converge uniformly. As shown in previous works \cite{Fitzpatrick:2014vua,Simmons-Duffin:2016wlq,Komargodski:2012ek,Lemos:2017vnx}, the region of the defect spectrum responsible for reproducing the pole in $1-\bar{z}$ is that of fixed transverse twist $2\hat{h}$
and large transverse spin $s$. A subtlety is that the large transverse spin $s$ limit and the lightcone $1-\bar{z}$ limit cannot be taken independently. Indeed, there is an interplay between these two limits which highlights the spin range responsible for reproducing the bulk OPE limit \cite{Lemos:2017vnx}
\begin{equation}
    s\sim \frac{1}{1-\bar{z}}\,,
    \label{eq:limitszb}
\end{equation}
which should be contrasted with the usual quadratic relation $l^{2}\sim 1/(1-\bar{z})$ that appears in the homogenous four-point function (here $l$ denotes the Lorentz spin in $SO(d)$).
This also holds in the bulk-bulk-defect three-point function that is the focus of this paper. Indeed, another way to see this behavior is through the Casimir-singularity trick \cite{Simmons-Duffin:2016wlq} as shown in Section \ref{section:Multipoint_dCFT}.

To perform the matching in the crossing equation \eqref{eq:crossing bb identity}, one therefore needs the large transverse spin expansion of the defect conformal block \eqref{eq:defect block bb}. Since the transverse spin appears in two distinguished hypergeometric functions, it is useful to isolate the different $s$-dependent pieces. Let us first consider the second hypergeometric function with argument $z\bar{z}$. For this function, we can apply a well-known identity of the ${}_{2}F_{1}$ hypergeometric functions\footnote{Another convenient way to perform an expansion of ${}_{2}F_{1}$ at large parameters is to use its integral representation and apply a saddle-point approximation. See, for example, Appendix A in \cite{Bissi:2024wur}.} analogous to the one used in \eqref{eq:trasf2F1a}

\begin{equation}
    {}_{2}F_{1}\left(c-a,b,c,\frac{z}{z-1}\right)=(1-z)^{b}\,{}_{2}F_{1}(a,b,c,z)\,,\quad z\notin(1,\infty),\,
\end{equation}
which allows us to shift the large-parameter $s$ dependence just in the third argument of the hypergeometric function. At this point, the hypergeometric function collapses to just a single term due to the large $s$ limit:

\begin{equation}
\begin{split}
    {}_2F_1\!\left(2\widehat{h}+s,\frac{p}{2},2\widehat{h}+s+1-\frac{p}{2}; z \bar z\right)&=(1-z\bar{z})^{-\frac{p}{2}}{}_2F_1\!\left(1-\frac{p}{2},\frac{p}{2},2\widehat{h}+s+1-\frac{p}{2}; \frac{z\bar{z}}{z\bar{z}-1}\right)\\
    &\overunderset{s\to\infty}{\bar{z}\to1}{\sim}(1-z)^{-\frac{p}{2}}\,.
\end{split}
\end{equation}
The other part of the defect conformal block we need to expand is the hypergeometric function that arises from the Gegenbauer polynomial $C_{s}^{(\frac{q-2}{2})}(\cos{\varphi})$, as a solution of the transverse Casimir equation. To this scope, it is important to keep track of the interplay \eqref{eq:limitszb} between the large transverse spin $s$ expansion and the lightcone limit $\bar{z}\rightarrow1$. Concretely, we take the coupled rescaling
\begin{equation}
    s\rightarrow\frac{s}{\epsilon}\,,\quad \bar{z}\to 1+ \epsilon (\bar{z}-1)\,, \quad \epsilon\rightarrow0\,.
\end{equation}
With this prescription, the hypergeometric function we are considering has the simple leading behavior 
\begin{equation}
    \lim_{\epsilon\rightarrow0} {}_{2}F_{1}\left( -\frac{s}{\epsilon},\frac{q}{2}-1,2-\frac{q}{2}-\frac{s}{\epsilon},\frac{z}{1+\epsilon(1-\bar{z})}\right)\sim(1-z)^{1-q/2}\,.
    \label{eq:large-s-Gegenbauer-hypergeo}
\end{equation}
Finally, in the lightcone limit $\bar{z}\to1$ and at large transverse spin, the defect-channel conformal block in \eqref{eq:defect block bb} becomes
\begin{equation}
    \hat{f}_{\hat{h},s}(z,\bar{z})\overunderset{s\to\infty}{\bar{z}\to1}{\sim}e^{-s(1-\bar{z})}z^{\hat{h}}(1-z)^{1-\frac{d}{2}}\,.
\end{equation}

\paragraph{Bulk-bulk-defect blocks in the defect channel}
Let us now consider the case of the bulk-bulk-defect three-point function. We want to study the large transverse spin behavior of the defect conformal blocks \eqref{eq:general-defect-blocks}. Since the conformal blocks factorize, it is once again more convenient to expand each term independently. Moreover, since the transverse part of the defect-channel blocks matches that of the bulk-bulk case, we can simply re-use~\eqref{eq:large-s-Gegenbauer-hypergeo} to extract its asymptotics.

We are left with the analysis of the asymptotic behavior of the Appell function $F_{4}$ \eqref{eq:appell F4} that appears in the defect-channel block.
In the lightcone limit $1-\bar{z}\rightarrow0$ the two variables $(v_{1},v_{2})$ remain finite, while all four parameters of $F_{4}$ scale with $s$ in the same way. For the large-$s$ expansion, it is convenient to use the reduction formula introduced in \eqref{eq:2F1 expansion}.
This representation expresses the Appell function as an infinite sum of products of two Gauss hypergeometric functions, and it is particularly useful for performing large-parameter asymptotic expansions.
In terms of the cross-ratios $(x,z,\bar{z})$, the variables $x_{1},x_{2}$ read
\begin{equation}
    x_{1}=\frac{(1+x^{2})z\bar{z}}{x^{2}(-1+z\bar{z})}\,,\quad x_{2}=\frac{x^{2}+z\bar{z}}{-1+z\bar{z}}\,.
\end{equation}
Since all Appell parameters scale in the same way at large transverse spin, it is convenient to separate their finite contributions:
\begin{equation}
    a=a^{(0)}+s\,, \quad b=b^{(0)}+s\,, \quad c_{1}=c_{1}^{(0)}+s\,, \quad \quad c_{2}=c_{2}^{(0)}+s\,.
\end{equation}
It is straightforward to see that the prefactor formed by the ratios of Pochhammer symbols is finite in the $s\rightarrow\infty$ limit
\begin{equation}
    \frac{(a^{(0)}+s)_{k}(b^{(0)}+s)_{k}(a^{(0)}+b^{(0)}-c_{1}^{(0)}-c_{2}^{(0)}+1)_{k}}{(c_{1}^{(0)}+s)_{k}(c_{2}^{(0)}+s)_{k}k!}\overset{s\to\infty}{\sim}\frac{(a^{(0)}+b^{(0)}-c_{1}^{(0)}-c_{2}^{(0)}+1)_{k}}{k!}\,.
    \label{eq:pochhammer exp}
\end{equation}
The two Gauss hypergeometric functions appearing on the RHS of \eqref{eq:2F1 expansion} have all their parameters large. Applying the following hypergeometric identity
\begin{equation}
    {}_{2}F_{1}(a,b,c;z)=(1-z)^{c-a-b} {}_{2}F_{1}(c-a,c-b,c,z)\,,
    \label{eq:transf2F1}
\end{equation}
we shift the large-parameter dependence into the prefactor $(1-z)^{c-a-b}$. In our case, the large parameter appears in the third argument of the hypergeometric function on the RHS, hence the hypergeometric function trivializes.
Combining this observation with the large-$s$ behavior of the Pochhammer prefactor in \eqref{eq:pochhammer exp}, we obtain the following leading asymptotic form of the Appell function $F_{4}$ in \eqref{eq:2F1 expansion}:
\begin{equation}
    \begin{split}
&\lim_{s\rightarrow\infty}F_{4}\left(a^{(0)}+s,b^{(0)}+s,c_{1}^{(0)}+s,c_{2}^{(0)}+s,x_{1}(1-x_{2}),x_{2}(1-x_{1})\right)\\
&\sim\sum_{k=0}^{\infty}\frac{(a^{(0)}+b^{(0)}-c_{1}^{(0)}-c_{2}^{(0)}+1)_{k}}{k!}\prod_{i=1}^{2}x_{i}^{k}(1-x_{i})^{c_{i}^{(0)}-a^{(0)}-b^{(0)}-s-k}\,.
    \end{split}
    \label{eq:appellexp}
\end{equation}

Combining this result with \eqref{eq:general-defect-blocks} and~\eqref{eq:large-s-Gegenbauer-hypergeo}, performing the sum with index $k$, and re-expressing variables in terms of the cross-ratios $(x,z,\bar{z})$, we obtain the leading behavior of the defect-channel conformal blocks in the simultaneous lightcone/large-spin limit:
\begin{equation}
    \hat{f}_{\hat{h}_{1},\hat{h}_{2},s}(z,\bar{z},x)\overunderset{s\rightarrow\infty}{\bar{z}\rightarrow1}{\simeq} e^{-s(1-\bar{z})}x^{-2(\hat{h}_{\hat\phi}-\hat{h}_{1}+\hat{h}_{2})}z^{\hat{h}_{2}}\left(\frac{(1+x^{2})(x^{2}+z)}{1+z}\right)^{\hat{h}_{\hat{\phi}}}(1-z)^{1-\frac{d}{2}}
\end{equation}
up to subleading corrections in the combined limit.

\section{Recasting the crossing equation as a series expansion}
\label{sec:solving-crossing}
The objective of this appendix is to explicitly perform the expansion of the LHS of~\eqref{eq:crossing_simplified} in the regime $z\ll x^2 \ll 1$
highlighted in Figure~\ref{fig:BBD-frame-limit}.

To begin with, we express the Lauricella function in the LHS of the crossing equation~\eqref{eq:crossing_simplified} via the Euler 
integral~\eqref{eq:Lauricella_def} and the change of integration variable to $v\equiv z+t(1-z)$, which runs from $z$ to $1$, yielding
\begin{equation}
    \begin{split}
        &\text{LHS}=\sum_{n}\lambda_{\Phi\Phi\mathcal{O}_\star}\, b^{(n)}_{\mathcal{O}_\star\hat\phi}\,(1-z)^{-\bar{h}_\star-2h_\Phi+\frac{d}{2}}
        z^{h_\Phi}x^{2\hat{h}_{\hat{\phi}}}\frac{\Gamma(2\bar{h}_\star)}{\Gamma(\bar{h}_\star)^{2}}\\[-3pt]
        &\hspace{140pt}\times \int_{z}^{1}\dd v\,(v-z)^{\bar{h}_\star-1}(1-v)^{\bar{h}_\star-1}\,\frac{v^{\hat{h}_{\hat{\phi}}-\bar{h}_\star}(x^{2}-v)^{2n}}{(x^{2}+v)^{2\hat{h}_{\hat{\phi}}+2n}}.
    \label{eq:new-crossing-Euler-v}
    \end{split}
\end{equation}
The next objective is to simplify the $v$ integral as much as possible and reduce it to a simpler hypergeometric form. 
This can be achieved by expanding the two $x$-dependent binomials, for which we use the simple expression
\begin{equation}
    (x^{2}-v)^{2n}=\sum_{j=0}^{2n}(-1)^{j}\binom{2n}{j}\,x^{2j}\,v^{2n-j}\,,
    \label{eq:finite-binomial}
\end{equation}
for the one with integer exponent, and the Mellin-Barnes representation
\begin{equation}
    \frac{1}{(x^{2}+v)^{2\hat{h}_{\hat{\phi}}+2n}}=\frac{1}{2\pi i}\int_{\mathcal{C}}\dd\sigma\,\frac{\Gamma(-\sigma)\Gamma(2\hat{h}_{\hat{\phi}}+2n+\sigma)}{\Gamma(2\hat{h}_{\hat{\phi}}+2n)}\,x^{2\sigma}\,v^{-2\hat{h}_{\hat{\phi}}-2n-\sigma}\,,
    \label{eq:MellinBarnes}
\end{equation}
for the one with non-integer exponent, where the contour $\mathcal{C}$ runs parallel to the imaginary axis within the range $-2\hat{h}_{\hat{\phi}}-2n<\mathrm{Re}\,\sigma<0$,
such that it separates the poles of the two Gamma functions in the numerator. The representation holds for any such contour; we will use this freedom below to fix the position of $\mathcal{C}$ more precisely, separately for each term of the sum over $j$.

The reason why we invoked a Mellin-Barnes representation instead of the simpler binomial expansion, is that the latter would only converge for either $v>x^2$ (if expanding in $x^2/v$)
or for $v<x^2$ (if expanding in $v/x^2$). The Mellin integral in expression~\eqref{eq:MellinBarnes}, instead, converges on the whole integration domain in $v$, and therefore allows 
for a swapping of the two integrals. 
Performing the swapping of the $v$ integral with both the $j$ sum and the Mellin integral, we get a Euler integral representation of the standard Gauss hypergeometric function
\begin{multline}
    \frac{\Gamma(2\bar{h}_\star)}{\Gamma(\bar{h}_\star)^{2}}(1-z)^{1-2\bar{h}_\star}\int_{z}^{1}\dd v\,(v-z)^{\bar{h}_\star-1}(1-v)^{\bar{h}_\star-1}v^{-\hat{h}_{\hat{\phi}}-\bar{h}_\star-j-\sigma}\\
    ={}_{2}F_{1}\!\left(\hat{h}_{\hat{\phi}}+\bar{h}_\star+j+\sigma,\bar{h}_\star;2\bar{h}_\star;1-z\right),
    \label{eq:v-integral-after-commutation}
\end{multline} 
which lets us recast~\eqref{eq:new-crossing-Euler-v} in the form
\begin{equation}
    \begin{aligned}
        &\sum_{n}\lambda_{\Phi\Phi\mathcal{O}_\star}\, b^{(n)}_{\mathcal{O}_\star\hat\phi}\,(1-z)^{\bar{h}_\star-2h_\Phi+\frac{d}{2}-1}z^{h_\Phi}\sum_{j=0}^{2n}(-1)^{j}\binom{2n}{j}\\
        &\hspace{25pt}\times\frac{1}{2\pi i}\int_{\mathcal{C}}\dd\sigma\,\frac{\Gamma(-\sigma)\Gamma(2\hat{h}_{\hat{\phi}}+2n+\sigma)}{\Gamma(2\hat{h}_{\hat{\phi}}+2n)}\,x^{2\hat{h}_{\hat{\phi}}+2j+2\sigma}\,{}_{2}F_{1}\!\left(\bar{h}_\star+\hat{h}_{\hat{\phi}}+j+\sigma,\bar{h}_\star;2\bar{h}_\star;1-z\right).
    \end{aligned}
    \label{eq:bulk-after-Mellin}
\end{equation}
In order to recast the hypergeometric function as an expansion in $z$, we employ the hypergeometric identity
\begin{equation}
    \begin{split}
        {}_{2}F_{1}\!\Bigl(\bar{h}_\star&+\hat{h}_{\hat{\phi}}+j+\sigma,\bar{h}_\star;2\bar{h}_\star;1-z\Bigr)\\
        &=\frac{\Gamma(2\bar{h}_\star)\Gamma(-\hat{h}_{\hat{\phi}}-j-\sigma)}{\Gamma(\bar{h}_\star)\Gamma(\bar{h}_\star-\hat{h}_{\hat{\phi}}-j-\sigma)}\,{}_{2}F_{1}\!\left(\bar{h}_\star+\hat{h}_{\hat{\phi}}+j+\sigma,\bar{h}_\star;1+\hat{h}_{\hat{\phi}}+j+\sigma;z\right)\\
        &+z^{-\hat{h}_{\hat{\phi}}-j-\sigma}\,\frac{\Gamma(2\bar{h}_\star)\Gamma(\hat{h}_{\hat{\phi}}+j+\sigma)}{\Gamma(\bar{h}_\star)\Gamma(\bar{h}_\star+\hat{h}_{\hat{\phi}}+j+\sigma)}\,{}_{2}F_{1}\!\left(\bar{h}_\star-\hat{h}_{\hat{\phi}}-j-\sigma,\bar{h}_\star;1-\hat{h}_{\hat{\phi}}-j-\sigma;z\right).
    \end{split}
    \label{eq:2F1-connection}
\end{equation}
When inserted in~\eqref{eq:bulk-after-Mellin}, this leads to two terms with different exponential behavior in the Mellin variable: the first term comes with the power $x^{2(\hat{h}_{\hat{\phi}}+j+\sigma)}$ and the second with $(x^{2}/z)^{\hat{h}_{\hat{\phi}}+j+\sigma}$.
This tells us that, in our regime of interest $z<x^2<1$, we can close the Mellin contour in $\sigma$ towards the right in the first term, and towards the left on the second one,
expressing this way the integral as sums over residues. 

The set of poles in the $\sigma$ plane of the two terms in~\eqref{eq:2F1-connection} is summarized in Table~\ref{tab:Mellin-poles}. 
\begin{table}[htp]
    \centering
    \renewcommand{\arraystretch}{1.3}
    \begin{tabular}{llcc}
        \hline
        Factor & Poles & G/S & E/NE \\
        \hline
        \multicolumn{4}{l}{\emph{First term of \eqref{eq:2F1-connection}, contour closed to the right}}\\
        $\Gamma(-\sigma)$ & $\sigma\in\mathbb{Z}_{\ge 0}$ & G & E \\
        $\Gamma(-\hat{h}_{\hat{\phi}}-j-\sigma)$ & $\hat{h}_{\hat{\phi}}+j+\sigma\in\mathbb{Z}_{\ge 0}$ & S & E \\
        ${}_{2}F_{1}(\dots;1+\hat{h}_{\hat{\phi}}+j+\sigma;z)$ & $\hat{h}_{\hat{\phi}}+j+\sigma\in \mathbb{Z}_{<0}$ & S & NE \\
        $\Gamma(2\hat{h}_{\hat{\phi}}+2n+\sigma)$ & $\sigma\in-2\hat{h}_{\hat{\phi}}-2n-\mathbb{Z}_{\ge 0}$ & G & NE \\
        \hline
        \multicolumn{4}{l}{\emph{Second term of \eqref{eq:2F1-connection}, contour closed to the left}}\\
        $\Gamma(2\hat{h}_{\hat{\phi}}+2n+\sigma)$ & $\sigma\in-2\hat{h}_{\hat{\phi}}-2n-\mathbb{Z}_{\ge 0}$ & G & E \\
        $\Gamma(\hat{h}_{\hat{\phi}}+j+\sigma)$ & $\hat{h}_{\hat{\phi}}+j+\sigma\in \mathbb{Z}_{<0}$ & S & E \\
        ${}_{2}F_{1}(\dots;1-\hat{h}_{\hat{\phi}}-j-\sigma;z)$ & $\hat{h}_{\hat{\phi}}+j+\sigma\in\mathbb{Z}_{>0}$ & S & NE \\
        $\Gamma(-\sigma)$ & $\sigma\in\mathbb{Z}_{\ge 0}$ & G & NE \\
        \hline
    \end{tabular}
    \caption{Simple poles of the two terms obtained by inserting \eqref{eq:2F1-connection} in \eqref{eq:bulk-after-Mellin}, with the factor they originate from and their location, for the contour $\mathcal{C}$ chosen as in \eqref{eq:contour-choice}. G (genuine) marks the poles of $\Gamma(-\sigma)$ and $\Gamma(2\hat{h}_{\hat{\phi}}+2n+\sigma)$, which are singularities of the full integrand; S (spurious) marks the poles at integer values of $\hat{h}_{\hat{\phi}}+j+\sigma$, which are present in each term separately but cancel in their sum. E (enclosed) and NE (not enclosed) refer to the closure of the term in question. Note that with the choice \eqref{eq:contour-choice} the pole at $\hat{h}_{\hat{\phi}}+j+\sigma=0$ lies to the right of $\mathcal{C}$, and is thus enclosed only by the closure of the first term.}
    \label{tab:Mellin-poles}
\end{table}
Among these are the singularities of the $\Gamma(-\sigma)\Gamma(2\hat{h}_{\hat{\phi}}+2n+\sigma)$ already present in~\eqref{eq:bulk-after-Mellin} and inherited by both terms, together with new ``spurious'' poles that are present in the individual terms but would cancel in the recombined expression. 
In order to simplify as much as possible the residue computation, let us recall that the contour $\mathcal{C}$ introduced in~\eqref{eq:MellinBarnes} need only be in the range $-2\hat{h}_{\hat{\phi}}-2n<\mathrm{Re}\,\sigma<0$, and that the Mellin integral of~\eqref{eq:MellinBarnes} commutes with the finite sum in $j$ of~\eqref{eq:finite-binomial}. 
This tells us both that the end result is independent of the relative location of the spurious poles with respect to the contour $\mathcal{C}$, and that this contour can be chosen independently for each $j$ term to lie within the range 
\begin{equation}
    \max\!\big(-2\hat h_{\hat\phi}-2n,\,-\hat h_{\hat\phi}-j-1\big)<\mathrm{Re}\,\sigma<-\hat h_{\hat\phi}-j\,.
    \label{eq:contour-choice}
\end{equation}
With this contour choice for each value of $j$, the anticipated contour closures only include poles of the $\Gamma(-\sigma)$ and $\Gamma(-\hat{h}_{\hat{\phi}}-j-\sigma)$ for the first term, and the poles of the $\Gamma(2\hat{h}_{\hat{\phi}}+2n+\sigma)$ and the $\Gamma(\hat{h}_{\hat{\phi}}+j+\sigma)$ for the second term, with the pole at $\hat{h}_{\hat{\phi}}+j+\sigma=0$ contributing only to the first term. The overall set of poles and the distinction on which ones are spurious and/or enclosed by the contour deformation is summarized in Table~\ref{tab:Mellin-poles}. We discuss the four different sets of enclosed poles in the list below, where ``first term'' and ``second term'' refer respectively to the first and second terms of~\eqref{eq:2F1-connection} inserted in~\eqref{eq:bulk-after-Mellin}.
\begin{enumerate}
    \item \label{firstpole} The poles of $\Gamma(-\sigma)$ at $\sigma=k \in \mathbb{Z}_{\ge 0}$ in the first term with a clockwise contour closure produce an overall residue equal to $(-1)^{k}/k!$ times the rest of the integrand evaluated at $\sigma=k$. These produce the term
    \begin{multline}
        \sum_{n}\lambda_{\Phi\Phi\mathcal{O}_\star}\, b^{(n)}_{\mathcal{O}_\star\hat\phi}\,(1-z)^{\bar{h}_\star-2h_\Phi+\frac{d}{2}-1}z^{h_\Phi}\frac{\Gamma(2\bar{h}_\star)}{\Gamma(\bar{h}_\star)}\sum_{j=0}^{2n}\sum_{k=0}^{\infty}(-1)^{j}\binom{2n}{j}\,\frac{(-1)^{k}(2\hat{h}_{\hat{\phi}}+2n)_{k}}{k!}\\
        \times \frac{\Gamma(-\hat{h}_{\hat{\phi}}-j-k)}{\Gamma(\bar{h}_\star-\hat{h}_{\hat{\phi}}-j-k)} x^{2\hat{h}_{\hat{\phi}}+2(j+k)}{}_{2}F_{1}\!\left(\bar{h}_\star+\hat{h}_{\hat{\phi}}+j+k,\bar{h}_\star;1+\hat{h}_{\hat{\phi}}+j+k;z\right).
    \end{multline}
    It is convenient to reorganize the sums in the last expression in order to just have one sum affect the exponent of $x$. We can then reorganize the two sums as a sum over $q=j+k \in \mathbb{Z}_{\ge 0}$ and a sum over $j$ with $0\le j \le \min (q,2n)$. This leads to the contribution
    \begin{equation}
        \begin{split}
        \sum_{n}\lambda_{\Phi\Phi\mathcal{O}_\star}\, b^{(n)}_{\mathcal{O}_\star\hat\phi}\,&(1-z)^{\bar{h}_\star-2h_\Phi+\frac{d}{2}-1}z^{h_\Phi}\frac{\Gamma(2\bar{h}_\star)}{\Gamma(\bar{h}_\star)}\sum_{q=0}^{\infty}\frac{\Gamma(-\hat{h}_{\hat{\phi}}-q)}{\Gamma(\bar{h}_\star-\hat{h}_{\hat{\phi}}-q)}\\
        &\hspace{50pt}\times c^{(n)}_{q}\;x^{2\hat{h}_{\hat{\phi}}+2q}\;{}_{2}F_{1}\!\left(\bar{h}_\star+\hat{h}_{\hat{\phi}}+q,\bar{h}_\star;1+\hat{h}_{\hat{\phi}}+q;z\right)
        \end{split}
        \label{eq:pole_first_term_I}
    \end{equation}
    where 
    \begin{equation}
        c^{(n)}_{q}\equiv(-1)^{q}\sum_{j=0}^{\min(q,2n)}\binom{2n}{j}\frac{(2\hat{h}_{\hat{\phi}}+2n)_{q-j}}{(q-j)!}\,.
        \tag{\ref{eq:c-coefficients}}
    \end{equation}
    \item The poles of $\Gamma(-\hat{h}_{\hat{\phi}}-j-\sigma)$ at $\sigma=m-j-\hat{h}_{\hat{\phi}}$ in the first term produce a residue analogous to the one above, namely equal to $(-1)^{m}/m!$ times the rest of the integrand evaluated at this point. This leads directly to
    \begin{equation}
        \begin{split}
        \sum_{n}\lambda_{\Phi\Phi\mathcal{O}_\star}\, b^{(n)}_{\mathcal{O}_\star\hat\phi}\,(1-z)^{\bar{h}_\star-2h_\Phi+\frac{d}{2}-1}&z^{h_\Phi}\frac{\Gamma(2\bar{h}_\star)}{\Gamma(\bar{h}_\star)^{2}}\sum_{m=0}^{\infty}\frac{(1-\bar{h}_\star)_{m}}{m!}\\
        &\times G^{(n)}_{m}\;x^{2m}\;{}_{2}F_{1}\!\left(\bar{h}_\star+m,\bar{h}_\star;1+m;z\right)
        \end{split}
    \label{eq:pole_first_term_II}
    \end{equation} 
    where
    \begin{equation}
        G^{(n)}_{m}=G^{(n)}_{-m}\equiv \sum_{j=0}^{2n}(-1)^{j}\binom{2n}{j}\frac{\Gamma(\hat{h}_{\hat{\phi}}-m+j)\Gamma(\hat{h}_{\hat{\phi}}+2n+m-j)}{\Gamma(2\hat{h}_{\hat{\phi}}+2n)}\,.
        \tag{\ref{eq:G-coefficients}}
    \end{equation}
    \item The residues of the poles of $\Gamma(2\hat{h}_{\hat{\phi}}+2n+\sigma)$ at $\sigma=-2\hat{h}_{\hat{\phi}}-2n-k$, $k\in\mathbb{Z}_{\ge 0}$ in the second term lead directly to a $(-1)^{k}/k!$ times the integrand evaluated at that point. The evaluation of this turns out to be identical to the one of item~\ref{firstpole} up to a replacement of $x^{2\hat{h}_{\hat{\phi}}+2q}$ with $(z/x^{2})^{\hat{h}_{\hat{\phi}}+q}$ in~\eqref{eq:pole_first_term_I}.
    \item \label{lastpole} The residues of the $\Gamma(\hat{h}_{\hat{\phi}}+j+\sigma)$ at $\sigma=-m-\hat{h}_{\hat{\phi}}-j$ with $m \ge 1$ in the second term lead to $(-1)^m/m!$ times the integrand evaluated at this point. Its evaluation leads to~\eqref{eq:pole_first_term_II} with sum restricted to $m\ge 1$ and $x^{2m}$ replaced by $(z/x^2)^{m}$.
\end{enumerate}
With these residues computed, the expansion of the LHS of the crossing equation~\eqref{eq:crossing_simplified} as powers of $z/x^2$ and $x^2$ is almost complete. 
Its form is now expressible as a sum over the residues of items~\ref{firstpole}--\ref{lastpole}, i.e.
\begin{equation}
    \begin{aligned}
        \sum_{n}&\lambda_{\Phi\Phi\mathcal{O}_\star}\, b^{(n)}_{\mathcal{O}_\star\hat\phi}\,(1-z)^{\bar{h}_\star-2h_\Phi+\frac{d}{2}-1}z^{h_\Phi}\frac{\Gamma(2\bar{h}_\star)}{\Gamma(\bar{h}_\star)^{2}}\Bigg\{G^{(n)}_{0}\,{}_{2}F_{1}\!\left(\bar{h}_\star,\bar{h}_\star;1;z\right)\\
        &+\sum_{m=1}^{\infty}\frac{(1-\bar{h}_\star)_{m}}{m!}\,G^{(n)}_{m}\left[x^{2m}+\left(\frac{z}{x^{2}}\right)^{m}\right]{}_{2}F_{1}\!\left(\bar{h}_\star+m,\bar{h}_\star;1+m;z\right)\\
        &+\Gamma(\bar{h}_\star)\!\sum_{q=0}^{\infty}\frac{\Gamma(-\hat{h}_{\hat{\phi}}-q)}{\Gamma(\bar{h}_\star-\hat{h}_{\hat{\phi}}-q)}\,c^{(n)}_{q}\!\left[x^{2\hat{h}_{\hat{\phi}}+2q}\!+\!\left(\frac{z}{x^{2}}\right)^{\hat{h}_{\hat{\phi}}+q}\right]\!{}_{2}F_{1}\!\left(\bar{h}_\star\!+\!\hat{h}_{\hat{\phi}}\!+\!q,\bar{h}_\star;1\!+\!\hat{h}_{\hat{\phi}}\!+\!q;z\right)\!\Bigg\}\\
        =&\sum_{\hat{h}_1,\hat{h}_2} C_{\widehat{\mathcal{O}}_1\widehat{\mathcal{O}}_2}\,\left(x^2\right)^{\hat{h}_{1}}\left(\frac{z}{x^2}\right)^{\hat{h}_{2}}\,.
    \end{aligned}
    \tag{\ref{eq:crossing-solved}}
\end{equation}
The only remaining step is to explicitly expand the binomial and hypergeometric functions as
\begin{equation}
    (1-z)^{\bar{h}_\star-2h_\Phi+\frac{d}{2}-1}\,{}_{2}F_{1}\!\left(\bar{h}_\star+\kappa,\bar{h}_\star;1+\kappa;z\right)=\sum_{l=0}^{\infty}P_{l}(\kappa)\,z^{l}
    \tag{\ref{eq:hypergeometric_and_binomial_expansion}}
\end{equation}
with
\begin{equation}
    P_{l}(\kappa)=\sum_{i=0}^{l}\frac{\left(1-\bar{h}_\star+2h_\Phi-\frac{d}{2}\right)_{i}}{i!}\,\frac{(\bar{h}_\star+\kappa)_{l-i}(\bar{h}_\star)_{l-i}}{(1+\kappa)_{l-i}\,(l-i)!}\,.
    \tag{\ref{eq:P_coefficients}}
\end{equation}
With this at hand, all the ingredients to solve the crossing equation~\eqref{eq:crossing_simplified} are explicitly available.

\newpage

\providecommand{\href}[2]{#2}\begingroup\raggedright\endgroup


\end{document}